\documentclass[]{revtex4-2}
\usepackage{graphicx}
\usepackage{multirow}
\usepackage{array}
\usepackage{amsmath}

\begin{document}

\title{Ion wake-mediated dust interactions under PK-4 conditions: a generalized and compact potential formulation}

\author{Diana Jimenez Marti\textsuperscript{1}, Benny Rodriguez
	Saenz\textsuperscript{1}, Peter Hartmann\textsuperscript{1, 2}, Evdokiya Kostadinova\textsuperscript{3}, Truell
	Hyde\textsuperscript{1}, Lorin Swint Matthews\textsuperscript{1}}

\affiliation{\textsuperscript{1}Center for Astrophysics, Space Physics and
	Engineering Research (CASPER), Baylor University, Waco, Texas 76798-7310, USA.}
\affiliation{\textsuperscript{2}Institute for Solid State Physics and Optics, Wigner Research Centre for Physics, 49, Budapest H-1525, Hungary.}
\affiliation{\textsuperscript{3}Physics Department, Auburn University, Auburn 36849, Alabama, USA.}

\date{\today}

\begin{abstract}
	Dusty plasmas, composed of electrons, ions, neutral particles, and charged
	dust grains, exhibit self-organization phenomena such as string-like
	structures observed in microgravity experiments. The formation of these
	structures is influenced by ion wakes generated by streaming ions under
	external electric fields, as well as by time-evolving plasma inhomogeneities
	such as ionization waves. Existing ion wake models, such as point charge and
	Gaussian-based representations, often rely on configuration-specific
	parameters, limiting their general applicability. In this work, we present a
	robust and general potential model for dust and ion wake systems under
	PK-4-like conditions. Using a small set of coefficients determined from
	molecular dynamics simulations, the model captures the potential
	distributions for multiple interparticle distances. Its application to test
	cases and implementation in a small scale dust dynamics simulation
	demonstrates its applicability to a wide range of dust arrangements beyond
	string-like configurations.
\end{abstract}

\maketitle

\section{Introduction.}

Dusty plasmas provide a platform to explore new physical phenomena connecting
different areas of physics. Dusty (complex) plasmas consist of a three-component
plasma environment--electrons, ions and neutral particles--with suspended dust
particles, typically ranging in size from nanometers to micrometers. When
immersed in a plasma, these particles acquire electric charge, primarily through
the collection of ions and electrons \cite{melzer2019physics}. Since electrons
have higher mobility, the net charge of the dust is predominantly negative.
Charged dust grains in a plasma environment behave as an additional plasma
constituent. Due to the very low charge-to-mass ratio of dust grains, their
dynamic timescales are much slower compared to other plasma species. In
addition, their macroscopic dimensions allow these particles to easily be
observed and imaged, turning them into a powerful tool to investigate the
dynamics of each particle and the system at the kinetic level.

One interesting feature is the self-organization of dust grains in ordered
structures, which has been widely reported in experiments \cite{mitic2013three},
\cite{pustylnik2020three} \cite{thomas1994plasma}, \cite{chu1994direct},
\cite{pieper1996experimental}. The PlasmaKristall-4 (PK-4) system onboard the
International Space Station (ISS) has been employed to carry out dusty plasma
experiments under microgravity conditions since 2014 \cite{fortov2005project}.
Experiments performed in this facility have shown the formation of dust
string-like structures parallel to the direction of the axial electric field
\cite{mitic2013three}, \cite{pustylnik2020three}. This electrorheological (ER)
behavior of dusty plasmas enhances their relevance as a system for theoretical
and experimental research \cite{ivlev2008}, \cite{ivlev2010}. ER fluids consist
of microparticles suspended in a nonconducting fluid which has a different
dielectric constant from that of the particles. In such systems, an external
electric field polarizes the particles, leading to a dipole-dipole coupling
between them. Due to the dipole-dipole interaction, particles organize into
strongly coupled structures such as crystals, sheets and chains. In ER dusty
plasmas, the polarization arises from the distortion of the Debye shielding
cloud surrounding the negatively charged dust particles. However, the specific
processes responsible for the formation of these structures are still not fully
known and constitute an active field of research.

Under the action of an external electric field, positive ions acquire a net
drift velocity in the direction of the electric field. The attractive electric
force of a negatively charged dust grain acting on the streaming ions leads to
the formation of an ion wake, region of enhanced ion density downstream of the
grain, along the ion streaming direction. The attractive electric interaction
between dust and ion wakes has been proposed as one of the mechanisms leading
the formation of dust filaments \cite{melzer1996structure}.

\begin{figure}[ht!]

	\centering
	\includegraphics[width=80mm]{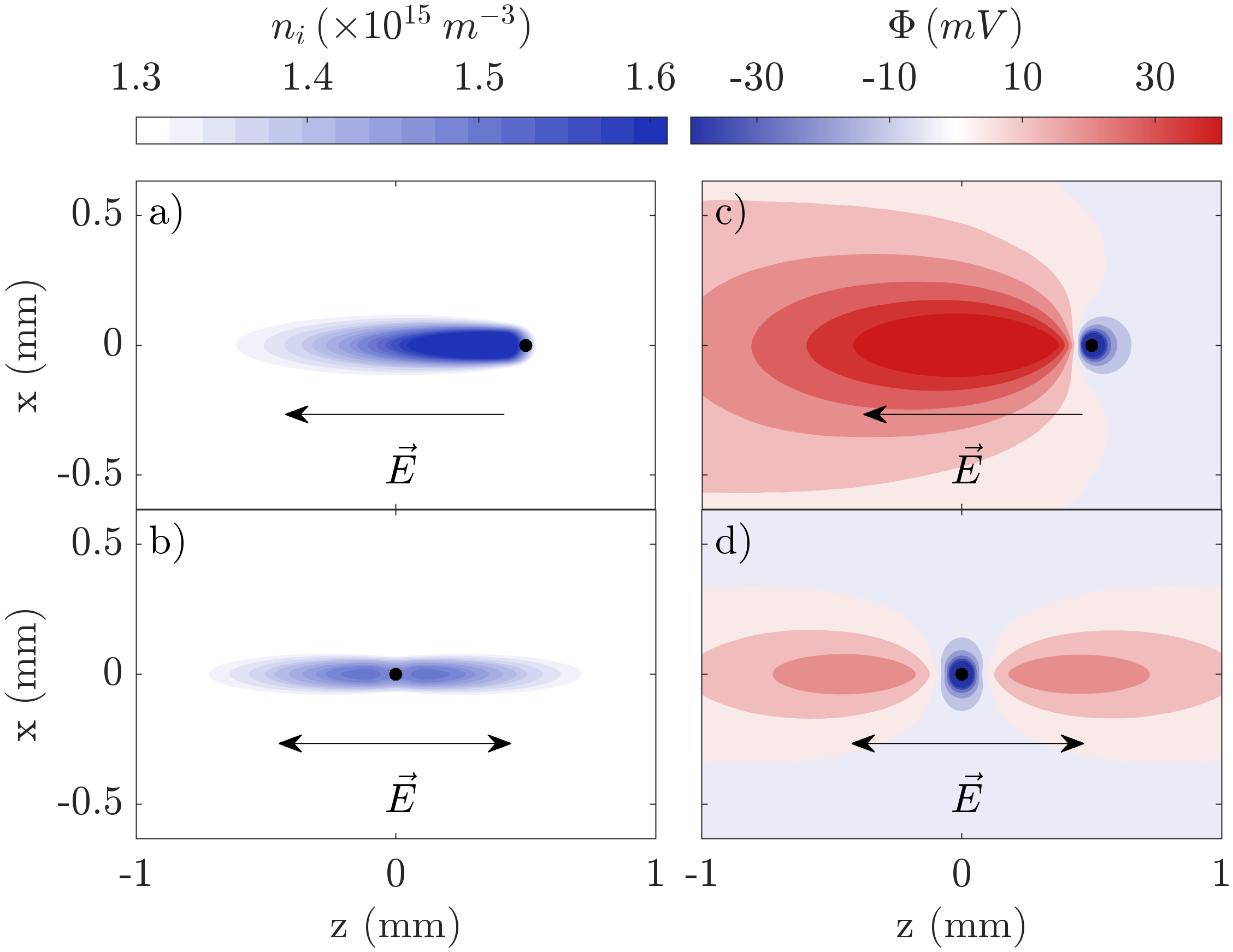}
	\caption{Representation of the ion density (a, b) and electric potential of a dust grain plus ion wake (c, d) for: a, c) a unidirectional electric field and b, d) a polarity-switching electric field. The positions of the dust grains are indicated by black dots.}
	\label{fig:ion_wake}

\end{figure}

The location of the ion wake depends on the type of experiment. In ground-based
experiments conducted in a Gaseous Electronic Conference radio-frequency (GEC
rf) cell, an external electric field frequency of 13.56 MHz is usually used
\cite{couedel2018full}, \cite{couedel2011wave}. Dust is levitated against
gravity by the electric field in the plasma sheath above the lower electrode.
Ions are accelerated from the bulk and flow towards the lower electrode. Hence,
the ion wake is formed below the dust grain (see Fig. \ref{fig:ion_wake}a, c). In
contrast, experiments under microgravity conditions such as the PK-4 DC plasma,
an alternating electric field is used to confine the dust grains in the field of
view with a typical frequency of 500 Hz \cite{mitic2013three}, \cite{ivlev2008}.
In this case, the ions can respond to the changing direction of the electric
field, leading to ion density enhancements on both sides of the dust grain (Fig.
\ref{fig:ion_wake}b, d).

The local plasma parameters in the PK-4 discharge, computed using a
two-dimensional Particle-In-Cell with Monte Carlo Collisions (PIC-MCC)
simulation, exhibit quasiperiodic moving inhomogeneities, known as ionization
waves \cite{hartmann2020ionization}. The fluctuations of the electric field
intensity and the electron and ion number densities in these waves can reach
amplitudes up to ten times their time-averaged value. Simulations presented in
Ref. \cite{Matthews2021}, in which dust charging and the coupled dynamics of
ions and dust grains were modeled, showed that the high electric field intensity
associated with the ionization waves are an important factor behind the
formation of the string-like dust structures observed in the PK-4 experiment.

Several models have been developed to describe the electric potential of the ion
wake, replacing explicit ion dynamics in large-scale simulations to reduce
computational cost while preserving the key physics of dust-wake interactions.
One of the simplest models is the so-called point charge model, in which the ion
cloud is replaced by a positive point charge with a symmetric Yukawa potential.
This approach has been widely used to investigate ground-based phenomena like
the non-reciprocal interaction in vertically aligned grains
\cite{schweigert1996alignment}, charging of interacting dust grains
\cite{matthews2020dust}, dynamics of 2D and 3D normal modes \cite{qiao2013mode},
\cite{melzer2014nonequilibrium}, among others. Other studies have been performed
using a modified point charge model by including also the dipole moment term in
the potential \cite{ishihara2000effect}, \cite{rocker2012mode}. In Ref.
\cite{kompaneets2007complex}, a kinetic approach is used to develop an
interaction potential model for dust particles in plasmas, describing the
potential of a point dust particle in a weakly ionized plasma with an ion drift
driven by an external electric field. A Gaussian representation of the ion wake
has also been used in several studies, which avoids the singularity that arises
near the ion wake region in the point charge model
\cite{vermillion2024interacting}, \cite{Mendoza2025}. In Ref.
\cite{vermillion2024interacting} this approach was used to identify minimum
energy configurations for two and three dust grains under ground-based
conditions. Each dust particle is described by three adjustable coefficients,
which are specific to the particle arrangement. Consequently, a different set of
coefficients is required for each configuration, making the application of these
models to dust dynamics simulations complex and time-consuming. In Ref.
\cite{Mendoza2025}, a Gaussian-based model was employed to describe the
differences in ion wakes formed under constant and time-evolving plasma
conditions due to ionization waves in the PK-4 experiment. In this formulation,
the number of coefficients also scales linearly with the number of dust grains
(in this case determined for a chain of four dust grains separated at $250 \,
	\mu m$) and are not transferable to other dust configurations.

In this work we present a more robust and general potential model formulation
for the electric potential distribution of the dust and ion wake system in the
vicinity of dust chains in representative PK-4 environments. It uses a small set
of coefficients that are shown to be extendable to a wide variety of dust
configurations, making it a versatile tool for complex plasma studies. We use a
molecular dynamics code to simulate the motion of ions to determine the charging
of dust and electric potential distribution near chains of dust grains under
time-evolving plasma conditions to account for the influence of ionization
waves. Four different interparticle distances are used to determine the fitting
coefficients as dust grains approach each other, with the aim of making the
model generalizable to dust arrangements beyond string-like configurations. The
model was implemented in a small-scale dust dynamics simulation to demonstrate
its practical applicability.

\section{Background.}

\subsection{PK-4 experiment.}

\begin{figure}[ht!]

	\centering
	\includegraphics[width=80mm]{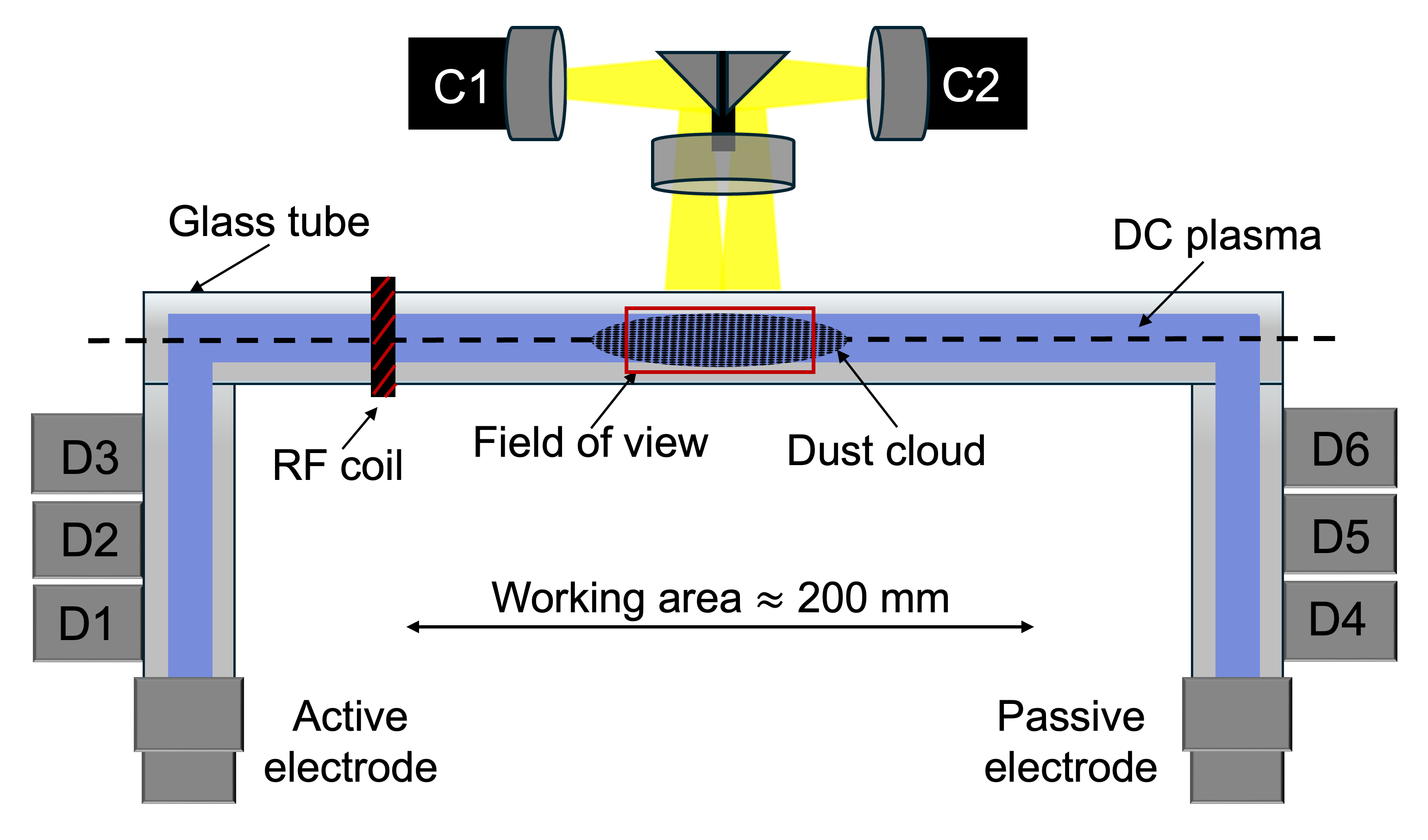}
	\caption{Scheme of the PK-4 experimental setup.}
	\label{fig:PK4_setup}

\end{figure}

The Plasmakristall-4 (PK-4) is a complex plasma experiment installed on the
International Space Station (ISS). A detailed description of this experiment can
be found in Refs. \cite{fortov2005project} and
\cite{pustylnik2016plasmakristall}. This microgravity experiment (see Fig.
\ref{fig:PK4_setup}) consists of a 30-mm diameter and 400-mm length $\pi$-shaped
glass tube. Two particle observation cameras (C1 and C2) are used to visualize
the dust. Three dust shakers on each side of the tube (D1 - D6) are used for
dispensing the dust grains into the tube. The active and passive electrodes are
located at the two ends of the tube. The experiment setup allows for the
generation a direct current (dc) or radiofrequency (rf) plasma discharge with
argon or neon gas and includes the option of polarity switching of the axial dc
electric field to trap the charged dust grains in the working area
(central region of the glass tube, 200-mm long). The polarity of the electrodes
is alternated at a high enough frequency ($f=500\,\text{Hz}$) that the grains
cannot respond to the change in the field direction (characteristic frequencies
of the dust dynamics are in the range 10-100 Hz \cite{fortov2005complex}).

\subsection{Formation of dust filaments.}

\begin{figure}[ht!]

	\centering
	\includegraphics[width=70mm]{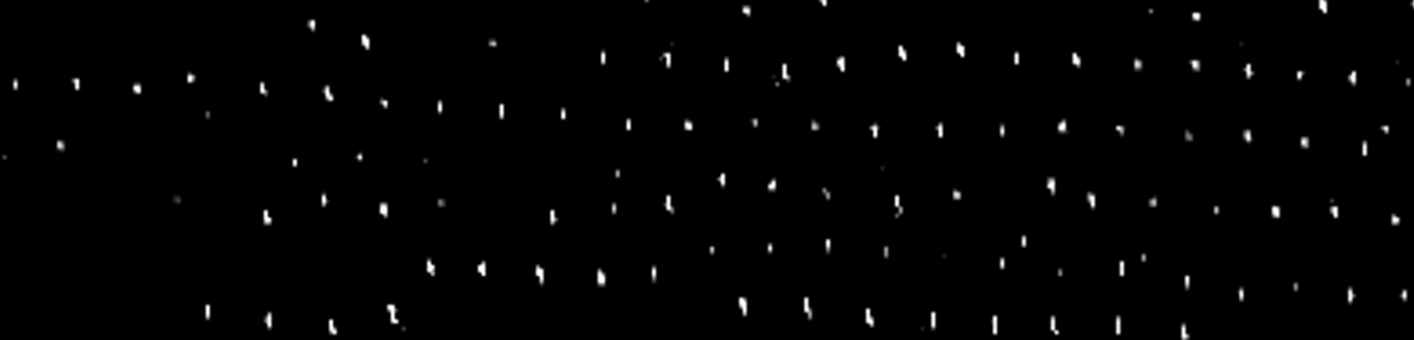}
	\caption{Dust filaments in the PK-4 experiment at 70.5 Pa neon gas pressure
		and 1 mA dc current \cite{gehr2025structural}.}
	\label{fig:filaments}

\end{figure}

It has been observed that dust grains under dc switching conditions can form
long and stable filaments along the direction of the axial alternating electric
field (see Fig. \ref{fig:filaments}) \cite{pustylnik2020three}. The average
interparticle distance within a filament is typically around $250-350 \, \mu m$
\cite{Matthews2021}, \cite{gehr2025structural}. The ions are light enough that
they can respond to the change in polarity of the electric field and flow
according to its direction. The ion trajectories are deflected by the negatively
charged dust grains and form a region of enhanced ion density known as an ion
wake (see Fig. \ref{fig:ion_wake}). In the case of a unidirectional electric field,
the ion density enhancement occurs behind the dust grain (Fig. \ref{fig:ion_wake}a,
c). For a polarity-switching electric field, ion focusing alternates its
direction, leading to an effective ion density enhancement on both sides of the
grain (Fig. \ref{fig:ion_wake}b, d). The interaction between ion wakes and dust
grains is the probable mechanism responsible for the formation of dust structures
\cite{melzer1996structure}, \cite{melzer1999transition}.

\subsection{Time-evolving plasma conditions.}

\begin{figure}[ht!]

	\centering
	\includegraphics[width=100mm]{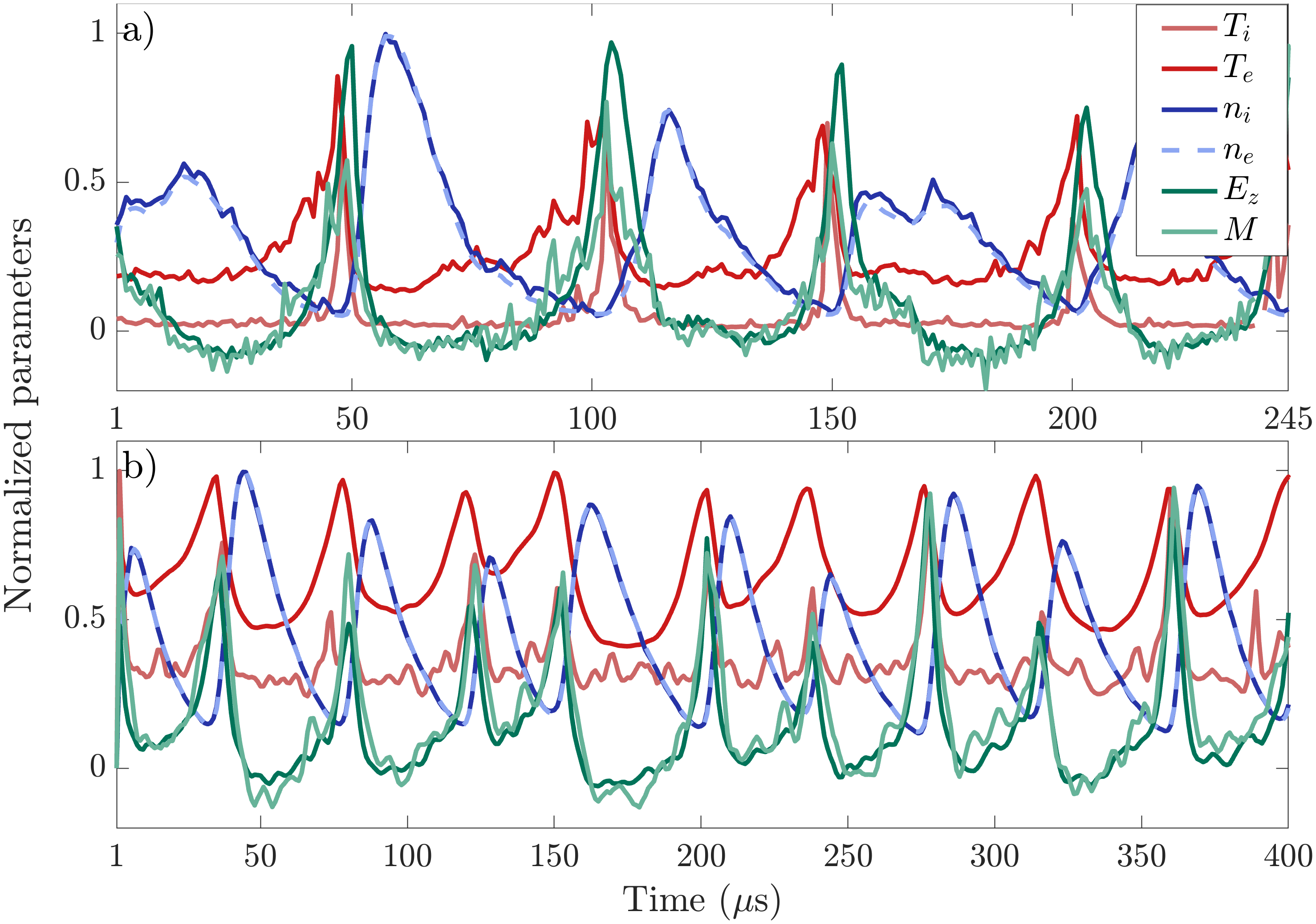}
	\caption{Representative segment of the time-evolving plasma parameters
		obtained from the PIC-MCC simulation at a) 40 Pa and b) 60 Pa. All
		magnitudes are normalized by their corresponding maxima.}
	\label{fig:plasma_params_complete}

\end{figure}

\begin{figure}[ht!]

	\centering
	\includegraphics[width=100mm]{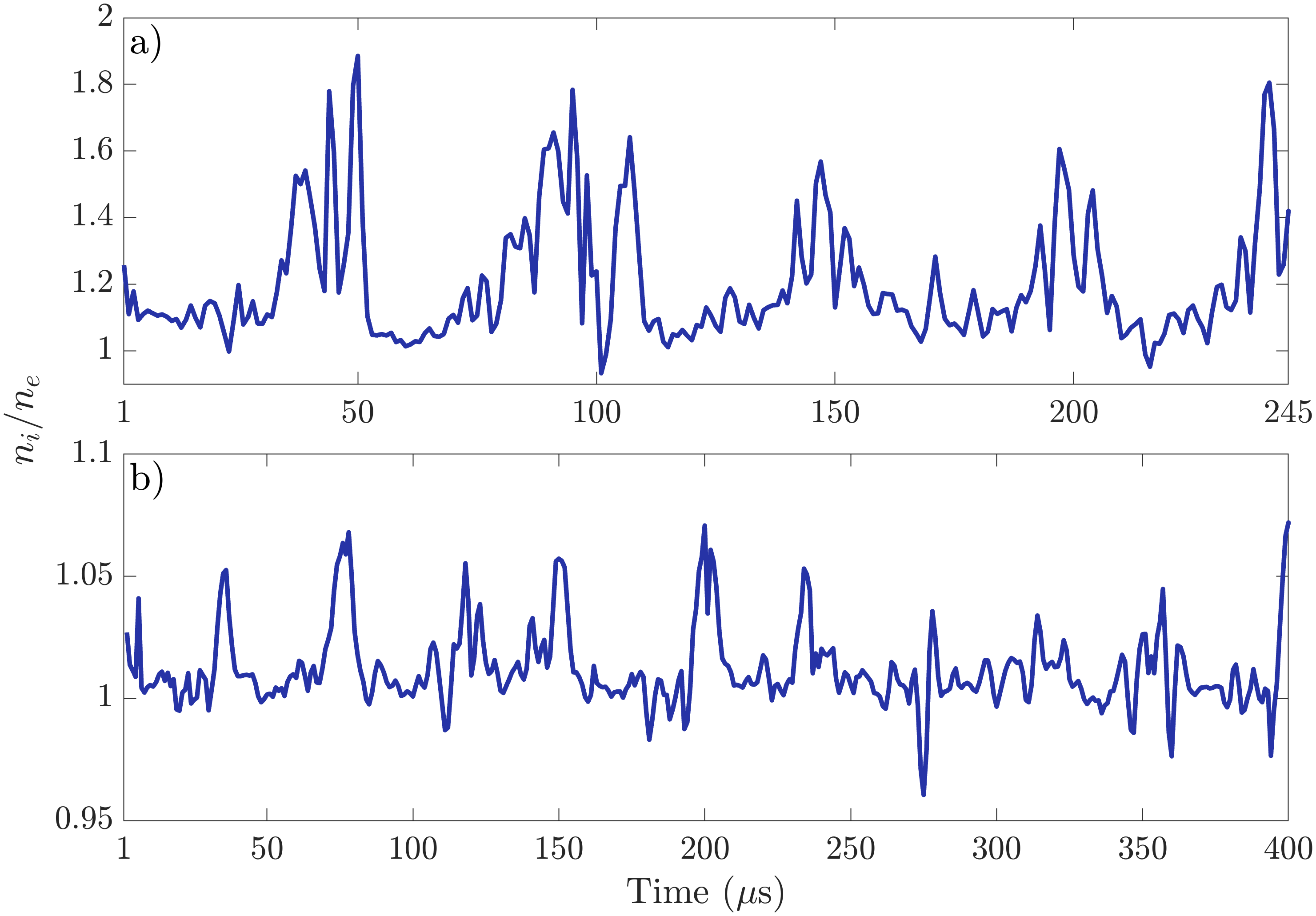}
	\caption{Ratio of ion to electron number density ($n_i/n_e$) obtained from
		the PIC-MCC simulation at a) 40 Pa and b) 60 Pa.}
	\label{fig:density_ratio}

\end{figure}

The local parameters in the PK-4 system are modeled with a 2D Particle-In-Cell
with Monte Carlo Collisions (PIC-MCC) simulation \cite{hartmann2020ionization}.
The gas temperature ($T_g$), discharge current (I) and dc voltage (V) used for
simulations at 40 Pa and 60 Pa neon gas pressures ($p$) are shown in Table
\ref{tab:sim_params_table} \cite{Matthews2021}, \cite{hartmann2020ionization}. This
model revealed the existence of quasiperiodic inhomogeneities moving along the
axial direction within the positive column of the PK-4 discharge, known as
ionization waves, which were confirmed by the ground based experiments in PK4-BU
\cite{hartmann2020ionization} (Fig. \ref{fig:plasma_params_complete}). Plasma
parameters such as electron temperature ($T_e$), ion temperature ($T_i$),
electron density ($n_e$) and ion density ($n_i$) were shown to vary by up to one
order of magnitude as an ionization wave passes a given position within the
discharge tube. The ratio between the ion and electron density ($n_i/n_e$)
obtained for the PIC-MCC simulation is shown in Fig. \ref{fig:density_ratio}. These waves have a frequency of the order of 10 kHz and a phase
velocity in the range of 500-1200 m/s. Although these timescales are much
faster than the dust response time, these temporal variations in the plasma
parameters have been shown to influence the dust ordering and ion wake features
\cite{Matthews2021}, \cite{Mendoza2025}.

\section{Methods.}

\subsection{Dynamic Response of Ions and Dust (DRIAD).}

To simulate the ions dynamics and dust charging, we use the N-body molecular
dynamics code DRIAD \cite{matthews2020dust}, which resolves the equations of
motion of these species according to their own timescales. The simulation does
not directly model the electrons, which are instead treated as a fluid governed
by a Boltzmann distribution that shields an ion's electric potential and
determines the electron current to dust grains. We use the simulation to
self-consistently obtain the charge on a fixed arrangement of dust grains and
the distribution of ions in the vicinity of the dust. To enhance computational
efficiency, we model the dynamics of super-ions, which are clusters of ions with
the same charge-to-mass ratio as a single ion. The number of ions per super-ion
is calculated as $n_iV_{sim}/N_{si}$, where $n_i$ is the ion number density,
$V_{sim}$ is the simulation volume and $N_{si}$ is the number of super-ions used
in the simulation. The super-ions that leave the simulation region or are
collected by a dust grain are re-injected with random positions and velocities on
the boundary consistent with the ion drift speed, following the method presented
in \cite{Hutchinson2002}.

The equation of motion for the ions (with mass $m_i$) is:

\begin{equation}
	m_i\ddot{\vec{r}}_i = \vec{F}_{id} + \vec{F}_{ii} + \vec{F}_{bound} + \vec{F}_z + \vec{F}_{in}.
	\label{eq:eq_of_motion_ions}
\end{equation}

The first term in the right-hand side represents the force on an ion $i$ with
charge $q_i$ due to dust grains with charge $Q_d$ located at a distance
$|\vec{r}_{id}|$ from the ion, which is described by a Coulomb interaction (Eq.
\ref{eq:ion_dust_force}), as electrons are depleted in the vicinity of the
negatively charged dust grain and the ions modeled in the simulation provide the
Debye shielding:

\begin{equation}
	\vec{F}_{id} = \sum_{d=1}^{N_d}\frac{q_iQ_d}{4\pi\epsilon_0}\frac{\vec{r}_{id}}{|\vec{r}_{id}|^3}.
	\label{eq:ion_dust_force}
\end{equation}

$\vec{F}_{ii}$ is the force from other ions inside the simulation and is given
by a screened Coulomb interaction to account for the screening provided by the
electrons:

\begin{equation}
	\vec{F}_{ii} = \sum_{j\neq i}^{N_i}\frac{q_iq_j}{4\pi\epsilon_0}(1
	+ \frac{|\vec{r}_{ij}|}{\lambda_{De}})exp{\left[-\frac{|\vec{r}_{ij}|}{\lambda_{De}}\right]}\frac{\vec{r}_{ij}}{|\vec{r}_{ij}|^3},
	\label{eq:ion_ion_force}
\end{equation}

\noindent where $q_{i, j}$ is the charge of the ions $i$, $j$, separated by the
distance $|\vec{r}_{ij}|$ and $\lambda_{De}=\sqrt{\epsilon_0k_BT_e/n_ee^2}$ is
the electron Debye length.

The confining electric force ${\vec{F}_{bound} = q_i\vec{E}_{bound}}$ arises
from the electric field $\vec{E}_{bound}$ due the ions outside the simulation
region. This field is derived from the gradient of the potential distribution
within an empty cylindrical cavity that matches the dimensions of the simulation
region immersed in a uniform plasma with density $n_o$, representing the density
far from the simulated region. The force $\vec{F}_z=q_i\vec{E}_z$ originates
from the interaction of the ions with the axial electric field $E_z\hat{z}$.
$\vec{F}_{in}$ is the force due to the ion-neutral collisions, which is
calculated using the null-collision method.

\begin{figure}[ht!]

	\centering
	\includegraphics[width=100mm]{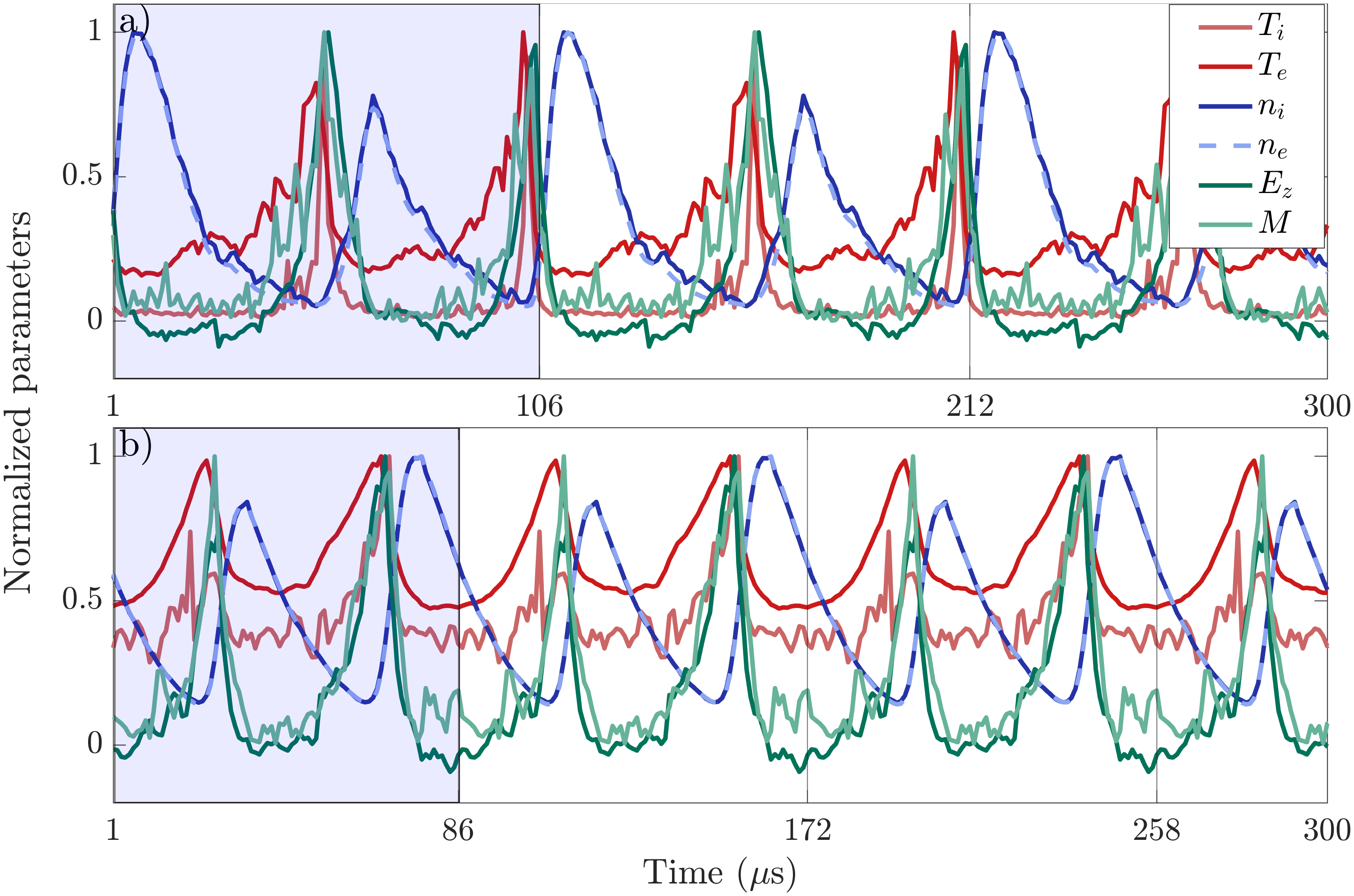}
	\caption{Representative segment of the time-evolving plasma parameters used
		in the simulation at a) 40 Pa and b) 60 Pa, covering a 300 $\mu s$ time
		interval. The selected portion from the PIC-MCC simulation (blue-shaded
		area) is repeated cyclically. All magnitudes are normalized by their
		corresponding maxima.}
	\label{fig:plasma_params}

\end{figure}

\begin{figure}[ht!]

	\centering
	\includegraphics[width=80mm]{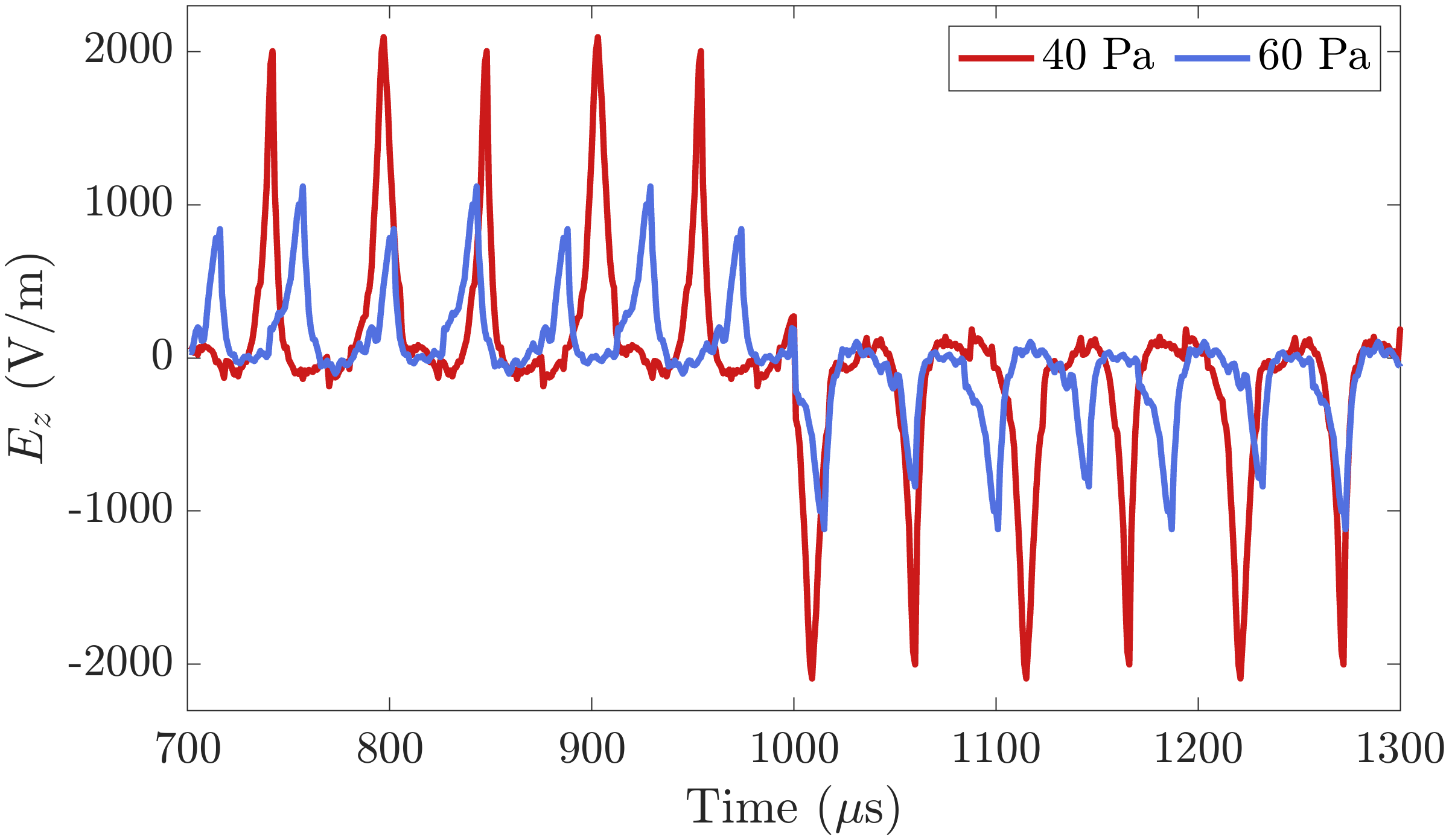}
	\caption{Representative segment of the time-evolving axial electric field
		($E_z$) used in the simulation covering a 600 $\mu s$ time interval. The
		polarity of the electric field switches at $t = 1000 \, \mu s$.}
	\label{fig:plasma_params_Ez_switching}

\end{figure}

In order to account for the effects of the time-evolving plasma conditions due
to the ionization waves, plasma conditions from a point at the center of the
PIC-MCC simulation are modeled, which includes data of the ion temperature
($T_i$), ion number density ($n_i$), electron temperature ($T_e$), electron
number density ($n_e$), axial electric field ($E_z$) and Mach number ($M$),
defined as $M=v_{i, z}/v_B$, where $v_{i, z}$ is the drift velocity of the ions
in the z-direction and $v_B = \sqrt{k_BT_e/m_i}$ is the Bohm velocity. A portion
of these evolving conditions (covering two ionization waves, blue-shaded areas
in Fig. \ref{fig:plasma_params}), are used to provide repeated boundary conditions.
The segment has a duration of 106 $\mu s$ in the 40 Pa case and 86 $\mu s$ for
60 Pa. Fig. \ref{fig:plasma_params_Ez_switching} shows a 600 $\mu s$ representative
segment of the axial electric field in its original units (non-normalized) for
both the 40 Pa and 60 Pa conditions. The direction of the electric field is
reversed every half-cycle ($1/2f=1\,ms$). The rest of plasma parameters are
repeated cyclically through the simulation. The time-average of each plasma
parameter over the selected portion of the PIC-MCC simulation are detailed in
Table \ref{tab:sim_params_table}. The average electron and ion Debye lengths ($\lambda_{De}$, $\lambda_{Di}$) are
calculated using the corresponding average number density and temperature.

\begin{table}[ht!]
	\begin{center}
		\caption{Parameters used in the PIC-MCC simulations of the PK-4
			experiment (Input parameters) and time-average plasma parameters
			over the selected portion (Time-average output parameters) for 40 Pa
			and 60 Pa gas pressures.}
		\label{tab:sim_params_table}
		\begin{tabular}{|c|c|c|}
			\hline
			Pressure (Pa)                  & 40    & 60           \\	  \hline
			\multicolumn{3}{|l|}{Input parameters}                \\	  \hline
			$T_g$ (K)                      & 300   & 300          \\	  \hline
			I (mA)                         & 0.8   & 2.0          \\	  \hline
			V (V)                          & 1000  & 770          \\      \hline
			\multicolumn{3}{|l|}{Time-averaged output parameters} \\	  \hline
			$n_{i} (\times 10^{15}m^{-3})$ & 0.219 & 1.26         \\	  \hline
			$n_{e} (\times 10^{15}m^{-3})$ & 0.197 & 1.25         \\	  \hline
			$T_i (\times 10^2 K)$          & 4.56  & 5.00         \\	  \hline
			$T_e (\times 10^2 K)$          & 737   & 499          \\	  \hline
			$\lambda_{Di} (\mu m)$         & 99.7  & 43.4         \\	  \hline
			$\lambda_{De} (\mu m)$         & 1336  & 435          \\	  \hline
			$E_z (V/m)$                    & 289   & 198          \\	  \hline
			$M$                            & 0.04  & 0.03         \\	  \hline
		\end{tabular}
	\end{center}
\end{table}

\subsection{Wake and dust potential model.}

The spatial distribution of the electric potential can be used to propose a
model for the potential of dust mediated by the ion wake in filamentary
structures. Because the ion wakes are aligned in the axial direction, the
electric potential distribution around the dust grains exhibits approximate
axial symmetry. Under these circumstances the variations in the azimuthal
direction are negligible compared with the radial and axial variations, making
any slice that contains a chain of dust grains representative of the full
three-dimensional structure. For this reason, the analysis focuses on data on
the $xz$ plane, where the most relevant features of the potential are fully
captured. Consequently, the model will be formulated directly in two dimensions,
which preserves all physically significant information while increasing the
stability of the fitting procedure.

\begin{figure}[ht!]

	\centering
	\includegraphics[width=85mm]{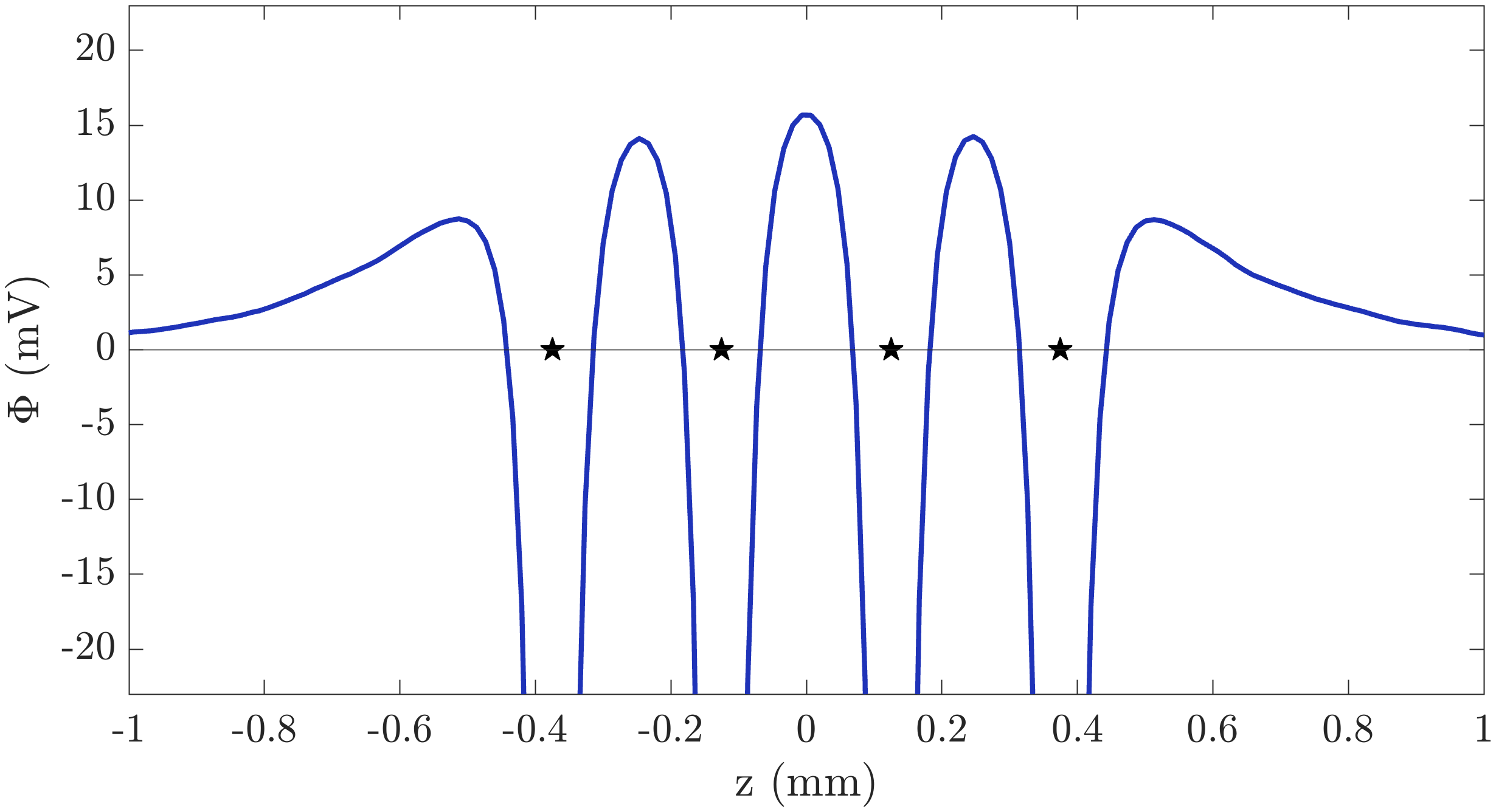}
	\caption{Electric potential at ($x=0, \, z$) for a chain of four dust grains
		located along the cylinder axis, separated by a distance $\Delta=250 \, \mu m$
		($p=60$ Pa, $T_g=300$ K, $V=770$ V) \cite{Mendoza2025}.
		Positions of the dust grains are indicated by black stars.}
	\label{fig:potential_example}

\end{figure}

Fig. \ref{fig:potential_example} shows a representative electric potential along the
$z$ axis for a chain of four dust grains located along the cylinder axis, under 60
Pa gas pressure, a discharge current of 2.0 mA and a dc voltage of 770 V
\cite{Mendoza2025}. The potential exhibits highly negative values near the dust
grain and increases rapidly towards zero as the distance from the dust grains
increases, interpreted as the presence of a term in the potential model that is
negative and inversely proportional to the distance from each grain. Between the
grains, one sees the presence of elevated electric potential (local maxima),
suggesting the presence of a positive contribution to the electric potential
from the ion wake. As the electric potential approaches zero at long distances,
both contributions should be decaying functions. Considering these aspects, we
propose a potential model, consisting of a screened Coulomb potential from each
grain and an anisotropic Gaussian function centered at each dust grain,

\begin{equation}
	\Phi(x, z)=\sum_{d=1}^{N_d}\frac{Q_d}{4\pi\epsilon_0}
	\frac{e^{-|\vec{r} - \vec{r}_d|/\lambda_d}}{|\vec{r} - \vec{r}_d|} +
	\alpha_d\frac{|Q_d|}{4\pi\epsilon_0}e^{-[\frac{x-x_d}{\sigma_{x, d}}]^2}e^{-[\frac{z-z_d}{\sigma_{z, d}}]^2},
	\label{eq:potential_model}
\end{equation}

\noindent where $Q_d$ and $\vec{r}_d = (x_d, z_d)$ are the charge and position
vector of the $d$-th dust grain. The coefficients $\lambda_d$, $\alpha_d$,
$\sigma_{x,d}$ and $\sigma_{z,d}$ are all positive. The coefficient $\lambda_d$
represents the isotropic shielding of the negative dust grain by the ions. A
small value of this coefficient represents a greater isotropic shielding. The
coefficient $\alpha_d$ represents the scaling factor for the positive
contribution to the wake potential. A greater value of this coefficient
indicates a stronger ion wake. The coefficients $\sigma_{x,d}$ and
$\sigma_{z,d}$ control the decay of this positive contribution in the $x$ and
$z$ directions respectively. A smaller value for these parameters represents a
shorter decay length. This potential model is a modified and more general
version of the model presented in \cite{Mendoza2025}. In the present case we
include the absolute value of the charge as a multiplicative factor that
controls the maximum of the Gaussian term along with the coefficient $\alpha_d$.
This takes into account that a higher dust charge implies that more ions will be
focused on the ion wake for a given external electric field, density and
pressure, resulting in a more prominent ion wake. The present formulation of the
model improves its generalizability, which will be addressed in Subsections
\ref{Data-driven model simplifications} and \ref{Potential model fitting}.

This model for the potential can be naturally extended to three dimensions (Eq.
(\ref{fig:potential_example})) by including a $y$-dependent term and an extra
coefficient $\sigma_{y, d}$ in the asymmetric Gaussian in Eq.
(\ref{fig:potential_example}). Coefficients $\sigma_{x,d}$ and $\sigma_{y, d}$ are
expected to be equal due to the symmetry of the system around the $z$ axis. The
two-dimensional case (Eq. (\ref{eq:potential_model})) is recovered when Eq.
(\ref{fig:potential_example}) is evaluated at $y = 0, \, y_d = 0$ ($xz$ plane).

\begin{equation}
	\Phi(x, y, z)=\sum_{d=1}^{N_d}\frac{Q_d}{4\pi\epsilon_0}
	\frac{e^{-|\vec{r} - \vec{r}_d|/\lambda_d}}{|\vec{r} - \vec{r}_d|} +
	\alpha_d\frac{|Q_d|}{4\pi\epsilon_0}e^{-[\frac{x-x_d}{\sigma_{x, d}}]^2}e^{-[\frac{y-y_d}{\sigma_{y, d}}]^2}e^{-[\frac{z-z_d}{\sigma_{z, d}}]^2}
	\label{eq:potential_model_3D}
\end{equation}

\section{Results.}

\subsection{Numerical modeling.}
\label{Numerical modeling}
To examine the electric potential distribution near a dust chain and determine
how the model coefficients evolve as dust grains approach each other, we
conducted simulations with DRIAD using chains of four melamine formaldehyde (MF)
spherical dust grains with radius $R_d$ = 1.69 $\mu m$ under 40 Pa and 60 Pa
neon gas pressures. The neutral gas temperature is 300 K. The dust mass is
$3.05\times10^{-14}$ $kg$ according to its radius and density (1510 $kg/m^3$). Four
different cases were examined, corresponding to the interparticle distances
$\Delta$ = 250 $\mu$m, 350 $\mu$m, 450 $\mu$m and 550 $\mu$m. The number of
super-ions considered in the simulations is 150 000 and 100 000, respectively.
The simulation geometry for both pressures is a cylinder of radius
2.0$\lambda_{De}$, while the length was set to 6.0$\lambda_{De}$ at 40 Pa and
12$\lambda_{De}$ at 60 Pa. These cylinder sizes and number of super-ions were
chosen to ensure that dust particles remain at least 2.0$\lambda_{De}$ away from
the simulation boundaries and that each super-ion has about 200 ions in both
cases. The ion time step is $\Delta t_i = 0.01$ $\mu s$. The representative
segments in the shaded areas shown in Fig. \ref{fig:plasma_params} from the PIC
simulation are used to account for the time-evolving boundary conditions in the
simulation for the two pressures. The evolving plasma parameters are updated
every $1 \, \mu s$ (100 ion time steps). The external electric field $E_z$, in
the axial direction, switches direction every 1 ms given $f=500 \, \text{Hz}$
with a 50$\%$ duty cycle. The total time simulated is 600 ms. The resulting ion
electric potential and dust charge are averaged over the last 580 ms after the
system is in equilibrium, covering 290 complete cycles of the switching electric
field.

\begin{figure}[ht!]

	\centering
	\includegraphics[width=140mm]{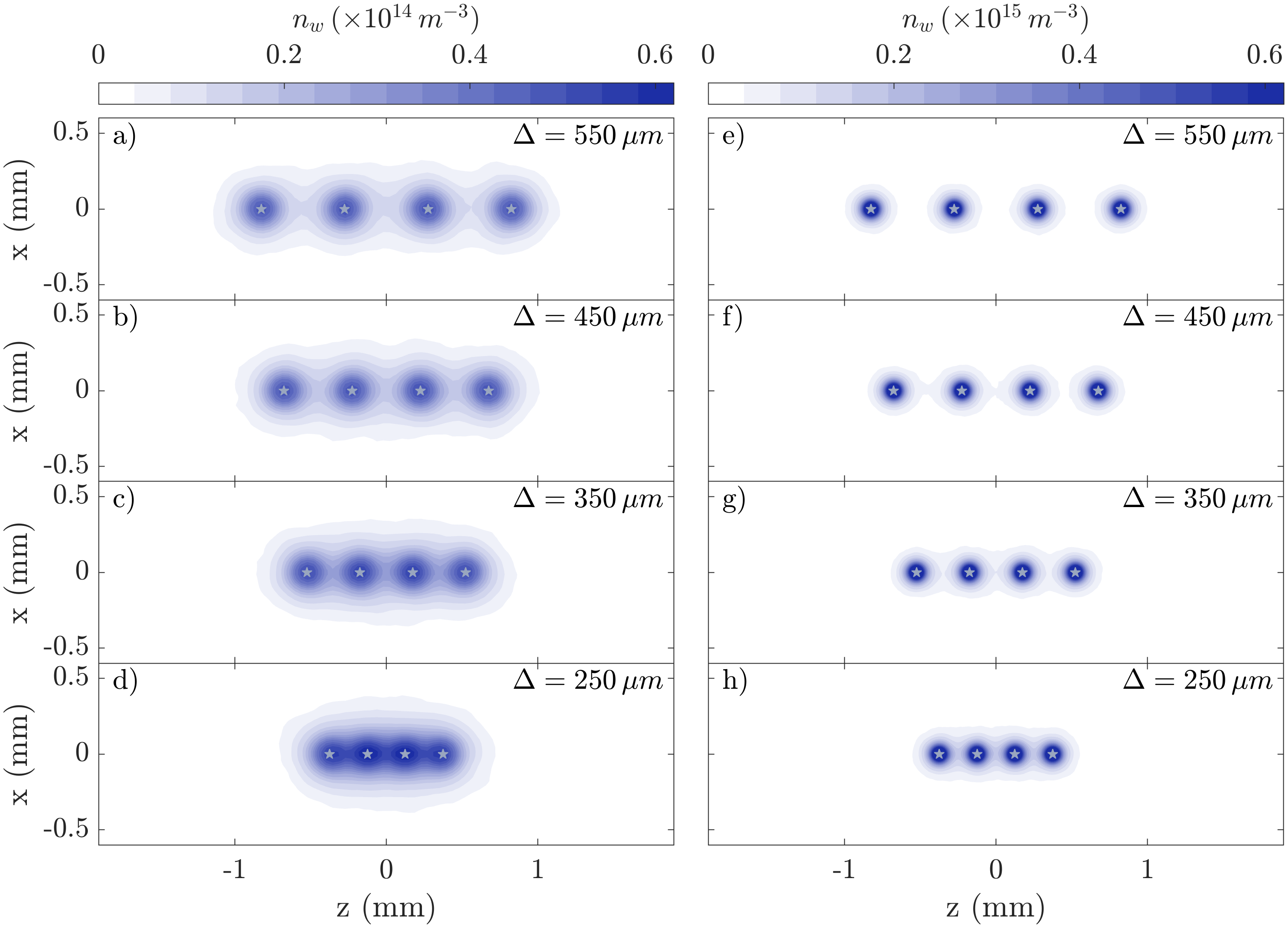}
	\caption{Ion density distribution of the ion wake obtained from DRIAD for: a-d) 40 Pa and
		e-h) 60 Pa gas pressure. The positions of the dust grains are
		represented by gray stars.}
	\label{fig:ion_density_data_plots}

\end{figure}

\begin{figure}[ht!]

	\centering
	\includegraphics[width=120mm]{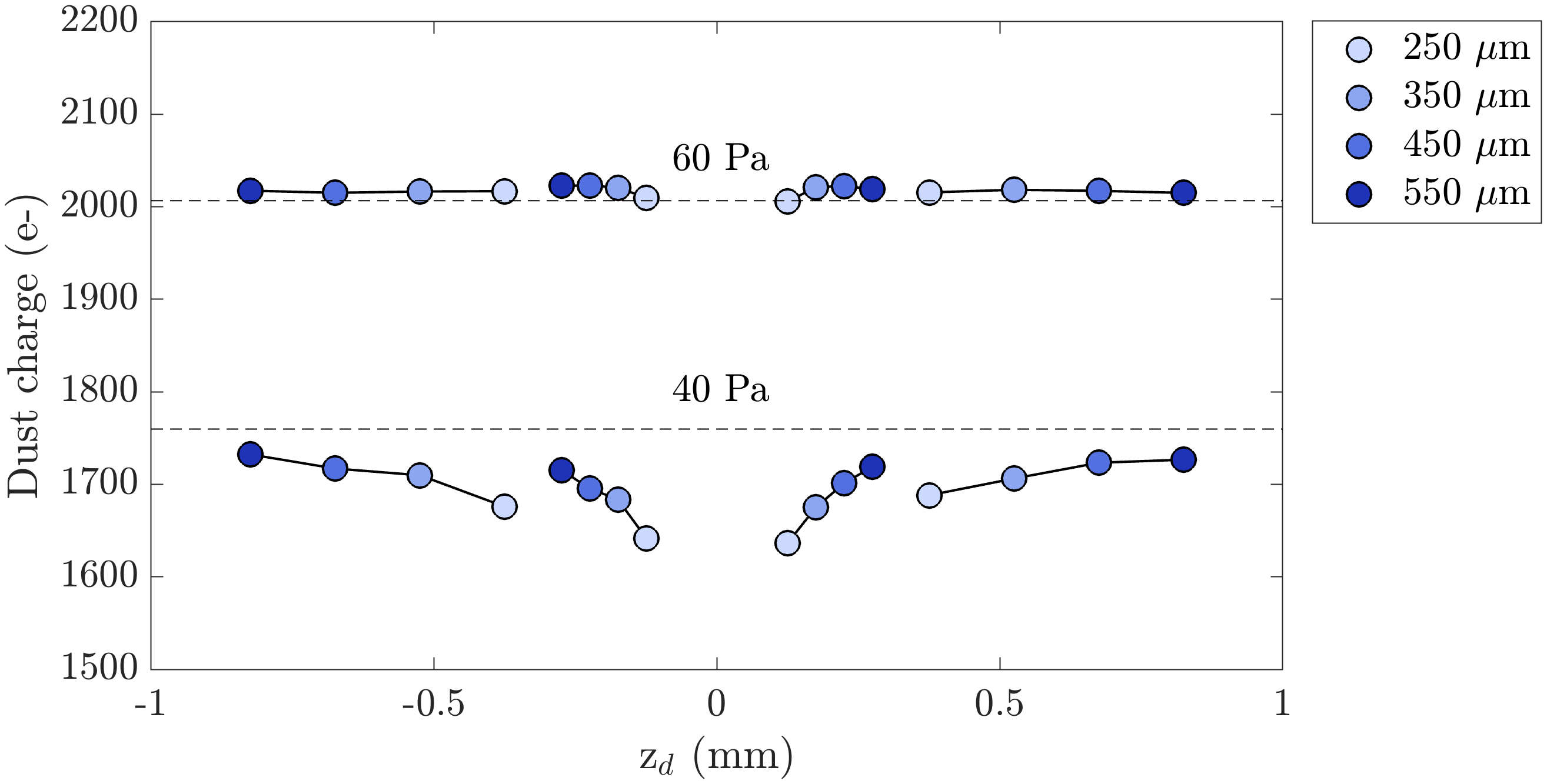}
	\caption{Charges of the four dust grains for 40 Pa and 60 Pa pressures as a
		function of the dust position. Marker colors indicate the different
		cases considered, corresponding to a specific interparticle distance.
		The horizontal dashed lines represent the charge obtained from
		simulations for a single dust grain at each pressure.}
	\label{fig:dust_charge}

\end{figure}

Fig. \ref{fig:ion_density_data_plots} shows the ion density distributions of the ion
wake from the simulations at each interparticle distance $\Delta$ under both
pressures. These distributions were calculated by subtracting the ion density of
the simulation without dust particles from that of each case, in order to
isolate the ion distribution caused by the presence of the dust. The average
charge of the dust grains at equilibrium is shown in Fig. \ref{fig:dust_charge} for
both pressures as a function of the dust position. The dust charge at 40 Pa is
lower than for 60 Pa, by about 15$\%$. In the 40 Pa case, ions have a slightly
lower temperature, making it easier to capture them. In addition, there is a
greater difference between the average electron and ion number densities $n_e$
and $n_i$ (see Fig. \ref{fig:density_ratio} and Table \ref{tab:sim_params_table}),
implying and enhanced ion current for this pressure. Together, these effects
increase the probability of capturing ions. In a given chain, the two inner dust
grains tend to collect less negative charge than the two outer dust grains, with
a more marked difference in the 40 Pa cases, since the ion current is enhanced
in this region. This difference increases as $\Delta$ decreases since the ion
density is enhanced as dust grains move closer (see Fig.
\ref{fig:ion_density_data_plots}). The dust charge tends towards that for an
isolated grain as $\Delta$ increases. This is an expected result since ions
approaching a given grain experience a reduced influence from neighboring
grains. As a result, each grain charges under conditions that are increasingly
independent and are closer to the single dust case.

\begin{figure}[ht!]

	\centering
	\includegraphics[width=140mm]{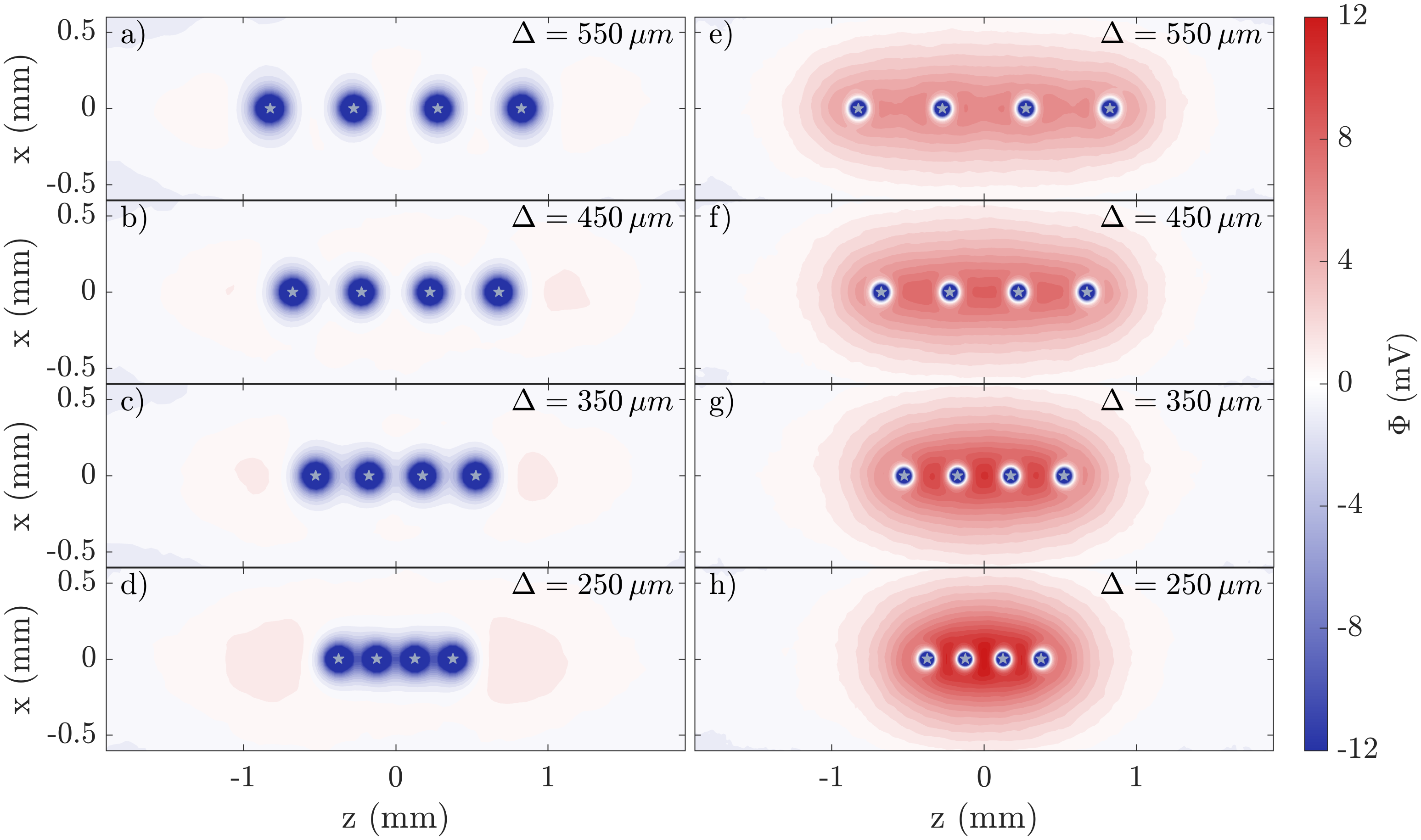}
	\caption{Potential distribution for the dust and ion wake system obtained
		from DRIAD for: a-d) 40 Pa and e-h) 60 Pa gas pressure. The positions of
		the dust grains are represented by gray stars.}
	\label{fig:potential_data_plots}

\end{figure}

The electric potential maps are shown in Fig. \ref{fig:potential_data_plots}. The
potential distributions show a highly negative potential in the region close to
the dust grains. In the 40 Pa cases (Fig. \ref{fig:potential_data_plots}, a-d) the
potential takes small positive values between the farthest spaced grains (Fig.
\ref{fig:potential_data_plots}, a) and becomes more negative as grains approach each
other (Fig. \ref{fig:potential_data_plots}, d). On the other hand, the potential in
the 60 Pa cases (Fig. \ref{fig:potential_data_plots}, e-h) reaches positive values
between dust grains and becomes more positive as $\Delta$ decreases. This is
consistent with a higher ion density in the 60 Pa case compared to the 40 Pa
case (see Table \ref{tab:sim_params_table}).

\section{Analysis and Discussion.}

\subsection{Data-driven model simplifications.}
\label{Data-driven model simplifications}

The potential model (\ref{eq:potential_model}) is formulated with independent
coefficients for each dust grain and for each interparticle distance, yielding a
total of 64 free coefficients (4 parameters per dust grain $\times$ 4 dust
grains $\times$ 4 configurations) for each pressure. The resulting coefficients
from preliminary fittings of this fully independent formulation to the data
showed no significant variation in the coefficients for grains within the same
chain and only minor differences between configurations. Moreover, no trend was
observed with the interparticle distance, i.e., the coefficients did not exhibit
a consistent increase or decrease as $\Delta$ changed. As shown in Fig.
\ref{fig:potential_data_plots}, the functional shape of the potential is essentially
the same in all cases for a given pressure. The only noticeable difference is a
gradual change in the ion wake intensity as the interparticle distance
decreases. The ion wake intensity is regulated by the dust charge (the higher
the absolute value of the charge, the more ions are attracted by the dust grain)
and by the distance between the grains, since wakes tend to overlap more
strongly as the particles move closer together. This dependence is already built
into the potential model, where the wake intensity scales proportionally with
charge and depends explicitly on the particle positions. The Mach number of the
ion flow also influences the wake intensity by controlling the degree of ion
focusing behind the grain. Since all four simulations considered for each
pressure were performed using the same plasma conditions, the flow regime
remains the same across cases for a given pressure. Such numerical behavior
suggests that the additional degrees of freedom might not correspond to real
physical differences, but rather to redundant parameters within the model.

Guided by these results, the model was progressively simplified in two steps.
First, the coefficients were assumed to be identical for all grains within a
chain ($\alpha_d=\alpha, \, \sigma_{x, d}=\sigma_x, \, \sigma_{z, d}=\sigma_z, \,
	\lambda_d=\lambda$). Second, the same set of coefficients was extended to all
cases for a given pressure.  This two-step simplification reduced the number of
free parameters from 64 to only 4 for each pressure, resulting in a more compact
formulation. This reduction greatly improves the numerical stability of the
fitting process while preserving the model's ability to capture the essential
features of the data. The resulting analytical form of the potential model after
these simplifications is:

\begin{equation}
	\Phi(x,z)=\sum_{d=1}^{N_d}\frac{Q_d}{4\pi\epsilon_0}\frac{e^
			{-|\vec{r} - \vec{r}_d|/\lambda}}{|\vec{r} - \vec{r}_d|} +
	\alpha\frac{|Q_d|}{4\pi\epsilon_0}e^{-[\frac{x-x_d}{\sigma_{x}}]^2}e^{-[\frac{z-z_d}{\sigma_{z}}]^2}.
	\label{eq:potential_model_simplified}
\end{equation}

\subsection{Potential model fitting.}
\label{Potential model fitting}

The positions of $N_d$ equally spaced dust grains in a chain arranged
symmetrically about the origin along the Z axis ($x_d = 0$, $y_d = 0$) can be
fully described by the interparticle distance instead of their four individual
positions. For practical implementation, the simplified expression of the
potential model Eq. (\ref{eq:potential_model_simplified}) was written in terms of
the interparticle distance $\Delta$ rather than the positions of the dust grains
as:

\begin{equation}
	\Phi(x,z)=\sum_{d=1}^{N_d}\frac{Q_d}{4\pi\epsilon_0}\frac{e^\frac{-\sqrt{x^2 +
				(z - k_d\Delta)^2}}{\lambda}}{\sqrt{x^2 + (z - k_d\Delta)^2}} +
	\alpha\frac{|Q_d|}{4\pi\epsilon_0}e^{-[\frac{x}{\sigma_x}]^2}
	e^{-[\frac{z - k_d\Delta}{\sigma_z}]^2}, \quad k_d = -\frac{2d-N_d-1}{2}
	\label{eq:potential_model_chain}
\end{equation}

This choice was made purely for computational simplicity, since expressing the
potential model as a function of $\Delta$ reduces the number of input parameters
required by the fitting routine. In order to determine the model coefficients, we
fit the expression Eq. (\ref{eq:potential_model_chain}) to the data obtained from
DRIAD (Fig. \ref{fig:potential_data_plots}). In general, it is important to prevent the model from overfitting the data, as this compromises its generalizability and degrades its performance when applied to different conditions \cite{mathworksoverfitting}. To prevent these limitations, we used the
\textit{Leave-p-Out Cross-Validation} (LpO CV) technique \cite{Brunton2021}. In
this method, the data (which includes $n$ samples) is split into a training set
and a validation set ($p$ samples). The training set ($n - p$ samples) is the
data that is used in the model fitting to determine the coefficients. The
validation set is not employed in obtaining the coefficients and is used just to
evaluate the quality of the fit. This procedure is repeated until all possible
subsets of size $p$ have been used for validation. As a result, multiple sets of
coefficients (specifically, $n!/(n-p)!p!$ sets) are obtained from the different
fitting cases. Here we define the resulting model coefficients as the set for
which the validation data is reproduced with the highest Coefficient of
Determination ($R^2$), which was calculated according to
\cite{mathworksRsquare}. The implementation of the LpO CV method supports the
model's generalizability to new dust configurations.

In this specific case, we used the $n=4$ data sets for each pressure to generate
all possible combinations of $p=2$ validation sets. The remaining $n-p=2$ sets for a
given combination were used as training sets. As a result, 6 sets of
coefficients are obtained from the different fitting cases. Exploratory fittings
were performed to identify the order of the coefficients. Based on this,
normalizations were applied to the potential expression to bring all
coefficients to the order of 1.0 and improve numerical stability during the
fitting. The original coefficients were then recovered by reversing the
normalizations.

\begin{figure}[ht!]

	\centering
	\includegraphics[width=140mm]{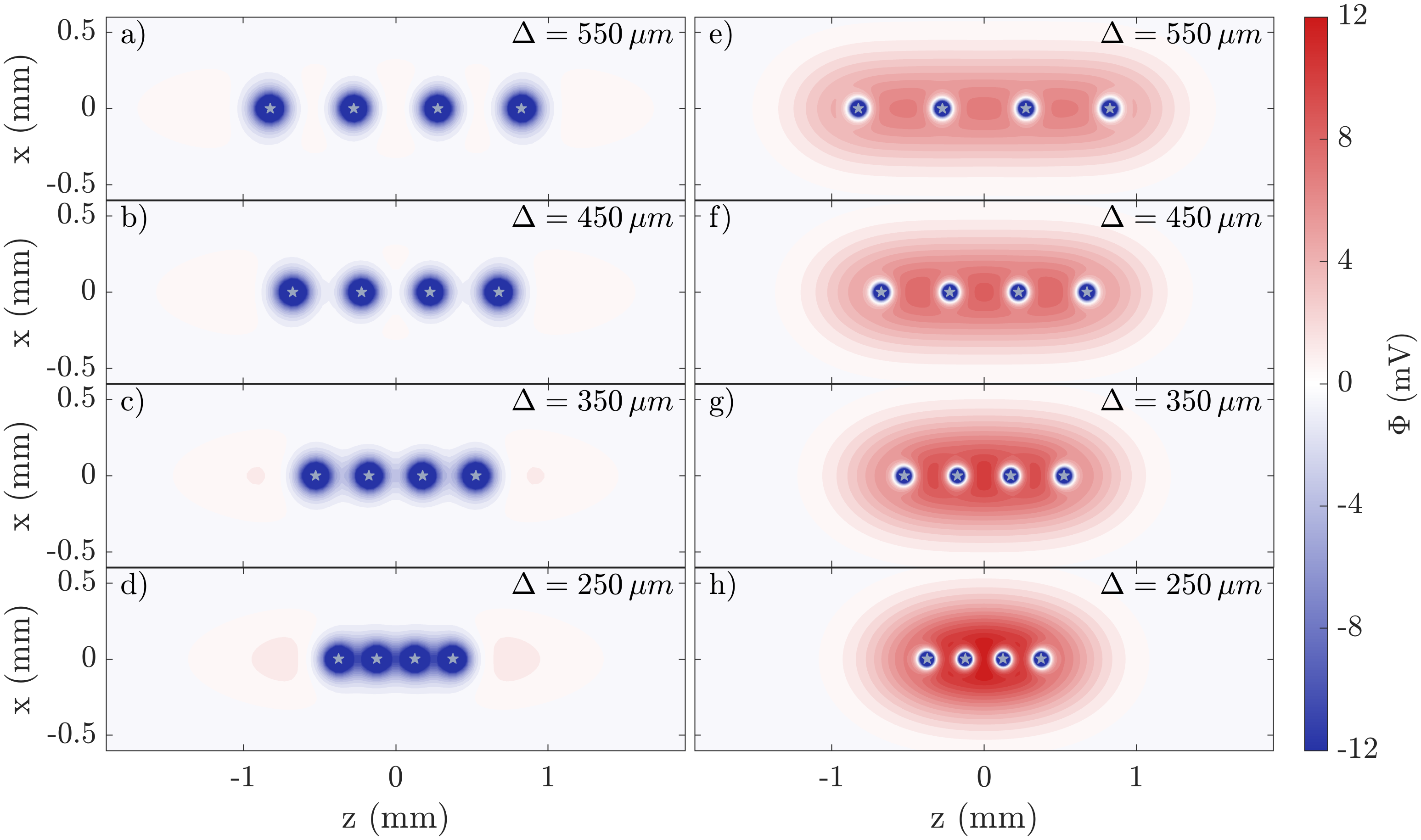}
	\caption{Potential distribution obtained with the potential model in Eq.
		(\ref{eq:potential_model_simplified}) for: a-d) 40 Pa and e-h) 60 Pa gas
		pressure. The positions of the dust grains are represented by gray
		stars.}
	\label{fig:potential_model_plot}

\end{figure}

The potential distributions obtained from the model Eq.
(\ref{eq:potential_model_simplified}) are shown in Fig. \ref{fig:potential_model_plot}.
The potential closely reproduces the simulation data in the entire simulation
region. These results appear smoother and do not exhibit the variations caused
by numerical noise observed in the simulated data.

\begin{figure}[ht!]

	\centering
	\includegraphics[width=140mm]{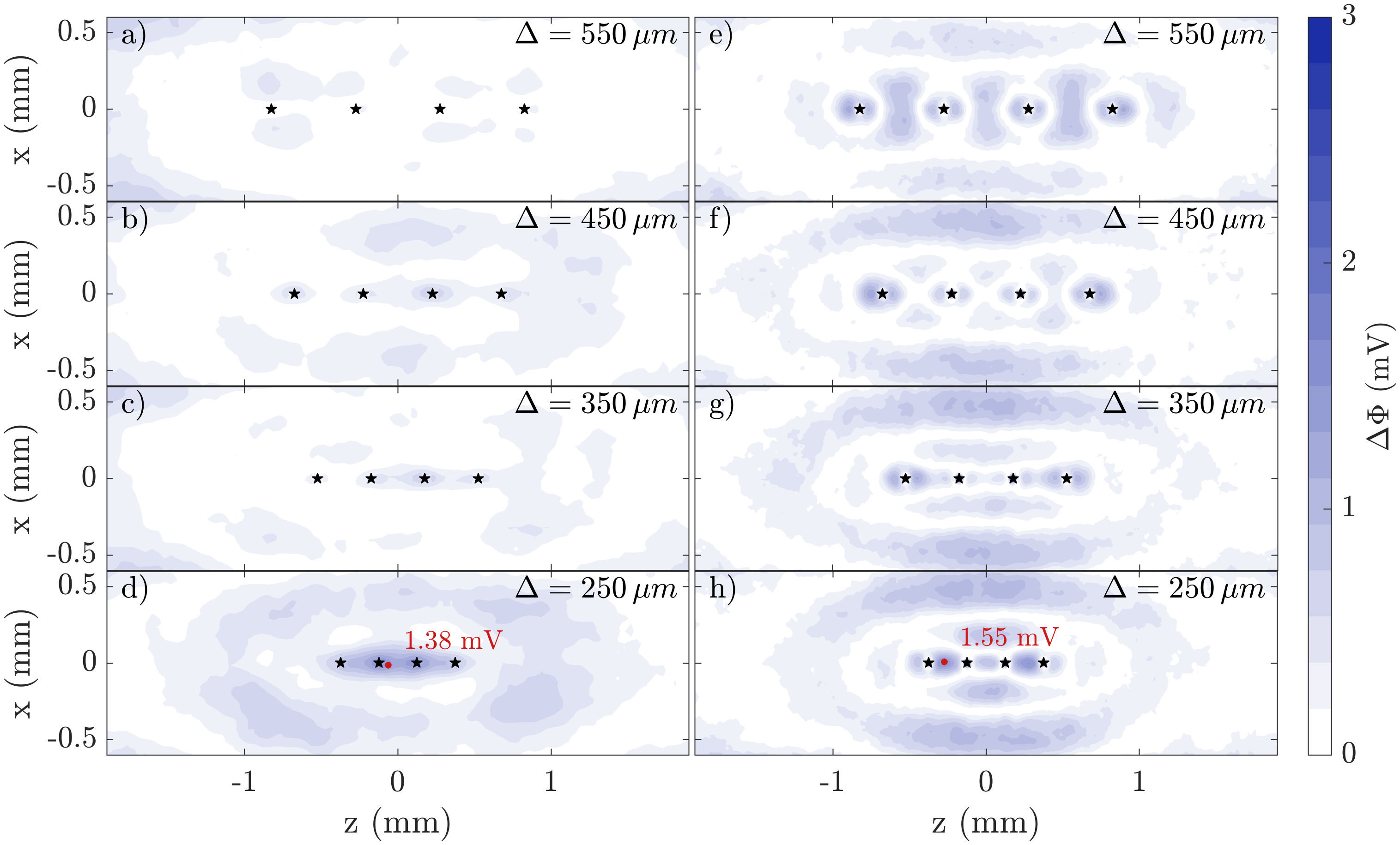}
	\caption{Absolute difference between the potential distribution obtained
		with the potential model in Eq. \ref{eq:potential_model_simplified} and the
		data obtained from DRIAD for: a-d) 40 Pa and e-h) 60 Pa gas pressure.
		The positions of the dust grains are indicated by black stars. The red
		dot indicates the position where the maximum difference is reached,
		while the adjacent red label shows its value.}
	\label{fig:potential_diff_plot}

\end{figure}

The absolute value of the differences between the potential distribution
predicted by the model Eq. (\ref{eq:potential_model_simplified}) and the data
obtained from the simulation (Fig. \ref{fig:potential_data_plots}) is shown in Fig.
\ref{fig:potential_diff_plot} for the eight cases under consideration. Table
\ref{tab:goodness_of_fit_table_train} shows the values of $R^2$ and Root Mean
Squared Error (RMSE). The difference plot indicates a good agreement between the
data and the model, as the maximum absolute differences remain below 2 mV in all
the cases. These differences are small when compared to the maximum absolute
value of the potential in the vicinity of the dust grains ($\sim$-31 mV for 40 Pa
and $\sim$12 mV for 60 Pa) and the dust surface potential (-1.4 V for 40 Pa and
-1.7 V for 60 Pa). The values of $R^2$ remain above $99\%$ and the RMSE is below
0.5 mV for all the cases. This suggests that the model accurately reproduces the
fundamental features of the data. The potential in the vicinity of the dust
grains is more accurately represented in the 40 Pa cases, as the difference is
less in this region. For both pressures, the absolute deviation between the
model and the simulation data increases as the interparticle distance decreases.
When particles are separated at longer distances, local errors around each dust
grain contribute independently to the absolute difference maps. As the
interparticle distance decreases, these errors spatially overlap, leading to a
higher discrepancy between the model and the data.

\begin{table}[h!]
	\centering
	\begin{tabular}{| c | c | c | c | c |}
		\hline

		\multirow{2}{*}{$\Delta \, (\mu m)$} & \multicolumn{2}{c|}{40 Pa} & \multicolumn{2}{c|}{60 Pa}
		\\ \cline{2-5}

		                                     & $R^2$ ($\%$)               & RMSE (mV)                  & $R^2$ ($\%$) & RMSE (mV) \\ \hline
		550                                  & 99.91                      & 0.212                      & 99.82        & 0.323     \\ \hline
		450                                  & 99.88                      & 0.238                      & 99.80        & 0.349     \\ \hline
		350                                  & 99.90                      & 0.214                      & 99.75        & 0.383     \\ \hline
		250                                  & 99.52                      & 0.386                      & 99.71        & 0.416     \\ \hline
	\end{tabular}

	\caption{Coefficient of Determination ($R^2$) and Root Mean Squared Error (RMSE) for the test cases.}
	\label{tab:goodness_of_fit_table_train}

\end{table}

Table \ref{tab:coefficients_table} presents the original coefficients obtained for
both pressures the ratio between them to compare the characteristics of the ion
wakes and evaluate how their properties vary for the two pressures. According to
Table. \ref{tab:coefficients_table}, the coefficient $\lambda$ is greater at 40 Pa
than at 60 Pa, implying a stronger symmetric shielding of the dust at 60 Pa
pressure. The scaling factor for the positive contribution $\alpha$ is greater
for the 60 Pa case than for the 40 Pa case, indicating a greater ion density at 60 Pa.
This agrees with the higher ion density and lower Mach obtained in the PIC-MCC
simulation for the higher pressure (see Table \ref{tab:sim_params_table}). The ratio
between the coefficients $\sigma_z$ and $\sigma_x$ quantifies the degree of
asymmetry of the ion wake and the direction of its elongation. For 60 Pa, the
value of $\sigma_z/\sigma_x$ is slightly greater than 1, suggesting an ion wake
that is approximately circular but moderately elongated along the axial
direction. At 40 Pa, this ratio is significantly greater than 1, revealing a
much more pronounced elongation. This behavior is consistent with the stronger
axial electric field in this case compared to the 60 Pa case (see Fig.
\ref{fig:plasma_params_Ez_switching}).

\begin{table}[ht!]
	\begin{center}
		\caption{Coefficients obtained from data fitting for 40 Pa and 60 Pa gas
			pressures, and the ratios of the values for the two cases
			($C_{40}/C_{60}$).}
		\label{tab:coefficients_table}
		\begin{tabular}{|c|c|c|c|}
			\hline
			                     & 40 Pa & 60 Pa.                            & $C_{40}/C_{60}$ \\
			\hline
			$\alpha$  ($m^{-1}$) & 400   & 1760                              & 0.227           \\	  \hline
			$\sigma_x$ ($\mu m$) & 284   & 337                               & 0.842           \\	  \hline
			$\sigma_z$ ($\mu m$) & 840   & 436                               & 1.93            \\	  \hline
			$\lambda$ ($\mu m$)  & 114   & 41.5                              & 2.76            \\	  \cline{1-4}
			$\sigma_z/\sigma_x$  & 2.96  & \multicolumn{1}{c}{1.29}   \vline                   \\	 \cline{1-3}
		\end{tabular}
	\end{center}
\end{table}

\subsection{Potential model evaluation on independent test cases.}
\label{Test cases}

\begin{figure}[ht!]

	\centering
	\includegraphics[width=150mm]{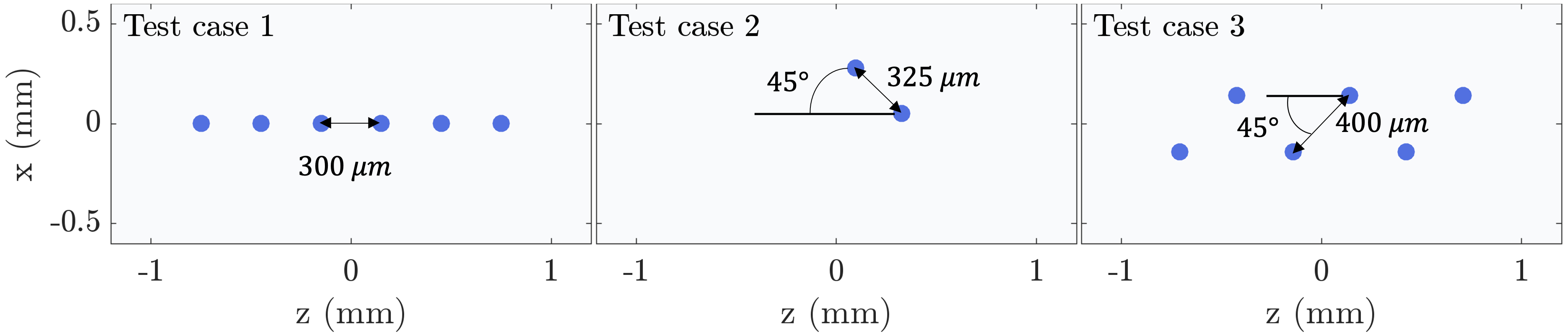}
	\caption{Schematic representation of the dust configurations used as test cases.}
	\label{fig:test_cases_dust_pos}

\end{figure}

We assess the generalizability of the potential model by applying it to test
cases. These additional simulations, which consist of dust configurations that
differ significantly from the ones used in determining the coefficients, allow
us to verify whether the model and the assumptions made remain valid for
different scenarios and can accurately reproduce the electric potential
distribution for a variety of relative dust positions. We used DRIAD to simulate
the cases shown in Fig. \ref{fig:test_cases_dust_pos}: a chain of six dust grains
with $\Delta=300$ $ \mu m$ (Test case 1), two dust grains separated by
$\Delta=325$ $ \mu m$ but at an angle to the axial electric field direction
(Test case 2) and a six-grain zigzag chain with $\Delta=400$ $ \mu m$ (Test case
3) with the same simulation parameters as the training cases.

\begin{figure}[ht!]

	\centering
	\includegraphics[width=140mm]{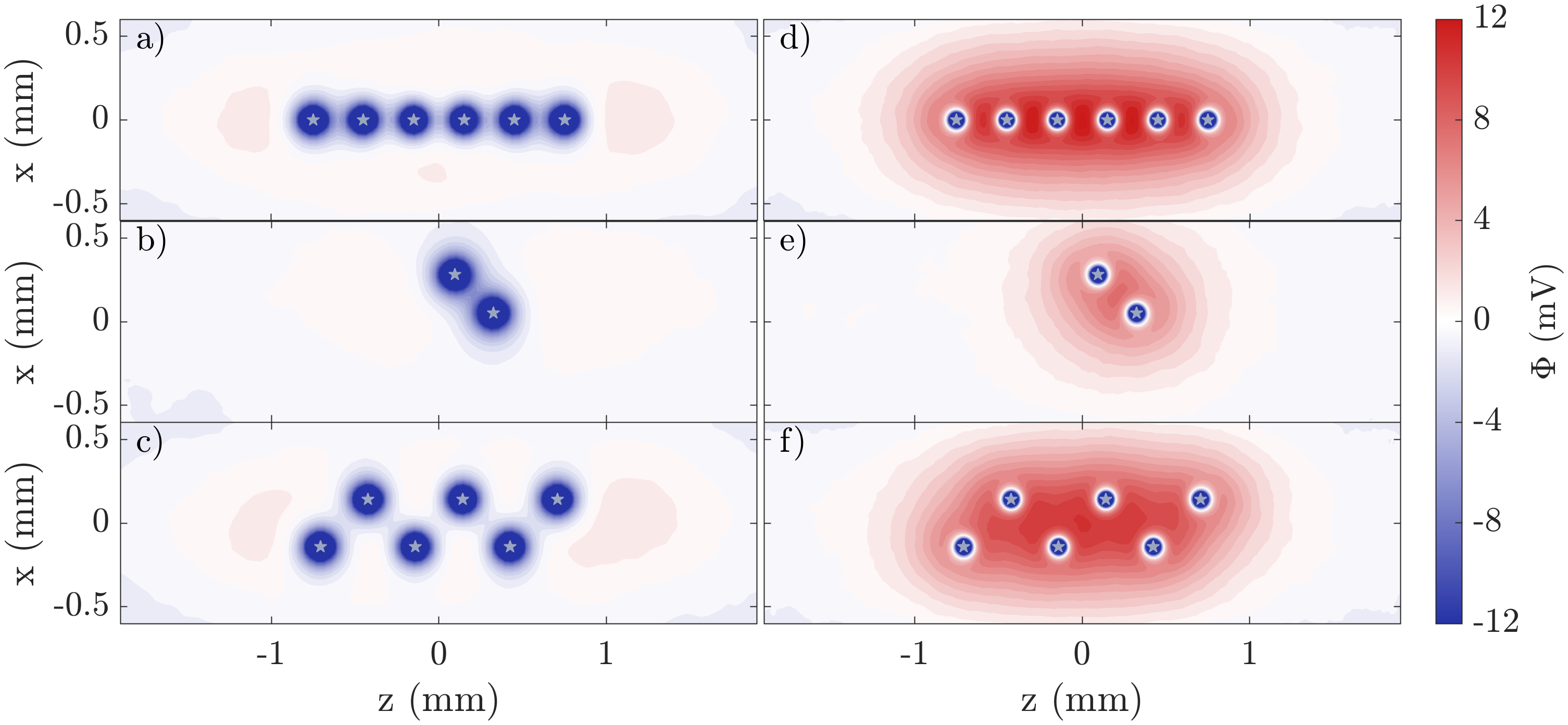}
	\caption{Electric potential distribution obtained from DRIAD for the test
		cases under: a-c) 40 Pa and d-f) 60 Pa gas pressures. The positions of
		the dust grains are indicated by gray stars.}
	\label{fig:test_data_plot}

\end{figure}

\begin{figure}[ht!]

	\centering
	\includegraphics[width=140mm]{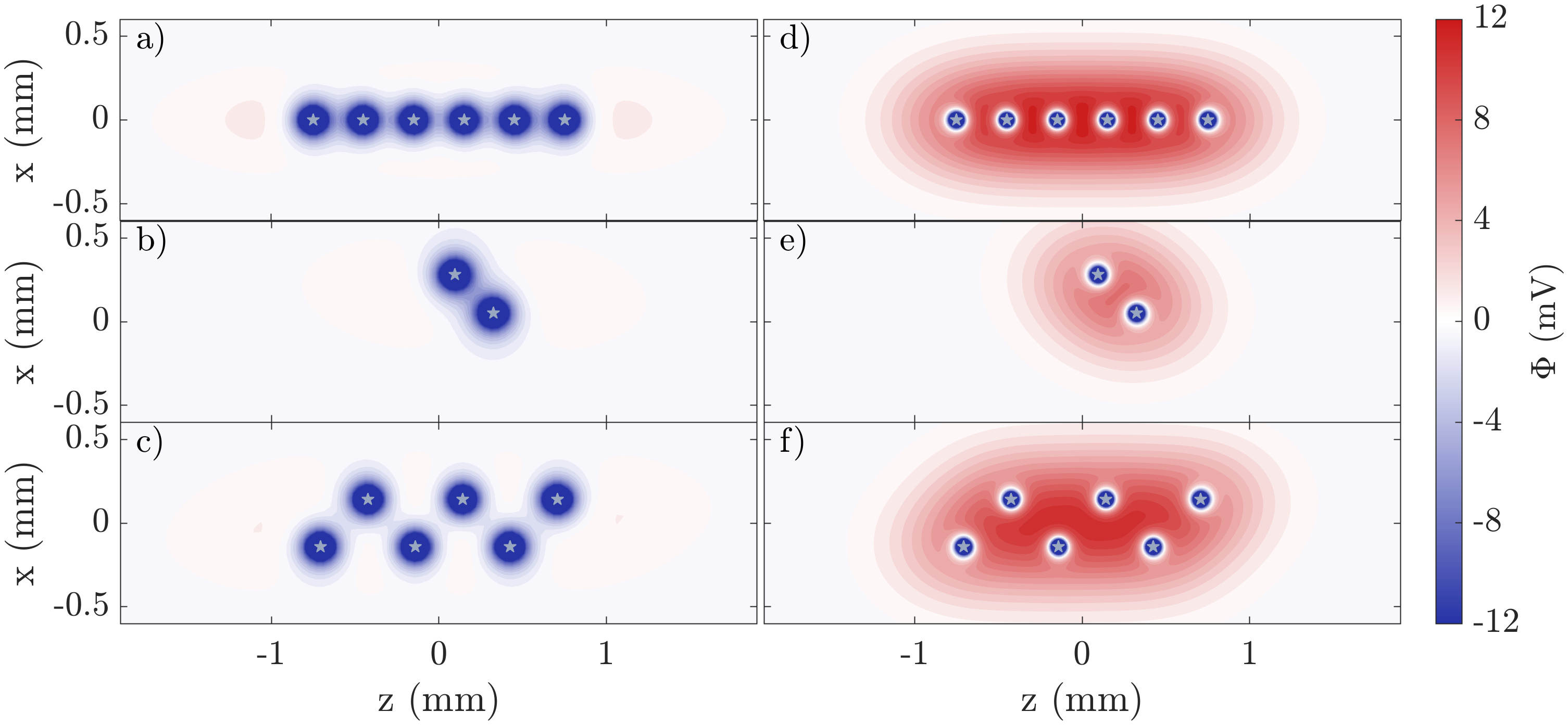}
	\caption{Electric potential distribution obtained with the potential model
		in Eq. \ref{eq:potential_model_simplified} for the test cases under: a-c)
		40 Pa and d-f) 60 Pa gas pressures. The positions of the dust grains are
		indicated by gray stars.}
	\label{fig:test_model_plot}

\end{figure}

\begin{figure}[ht!]

	\centering
	\includegraphics[width=140mm]{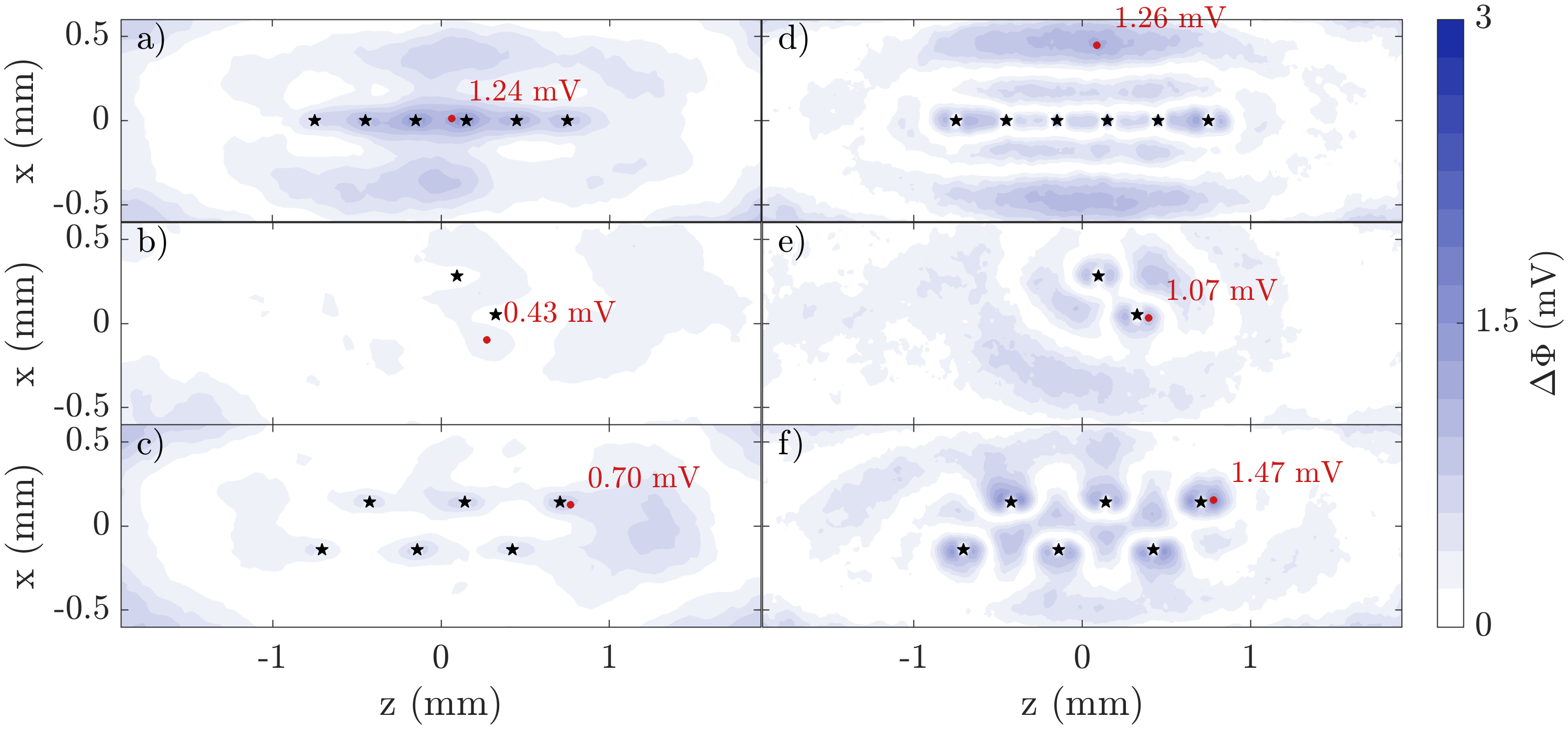}
	\caption{Absolute difference between the potential distribution obtained
		with the potential model in Eq. \ref{eq:potential_model_simplified} and the
		data obtained from DRIAD for the test cases for: a-c) 40 Pa and d-f) 60
		Pa gas pressures. The positions of the dust grains are represented by
		black stars. The red dot indicates the position where the maximum
		difference is reached, while the adjacent red label shows its value.}
	\label{fig:test_diff_plot}

\end{figure}

The electric potential distributions obtained from DRIAD for the test cases are
presented in Fig. \ref{fig:test_data_plot} for both pressures. The potential maps
generated by the model Eq. (\ref{eq:potential_model_simplified}), using the coefficients from Table \ref{tab:coefficients_table} and the dust charges predicted by DRIAD, are displayed in Fig. \ref{fig:test_model_plot}.
The potential model, reproduces the main features of the electric potential
distribution with good accuracy for all the test cases, with differences below
2.0 mV (see Fig. \ref{fig:test_diff_plot}), confirming that the simplified model closely matches the simulation
data. Table \ref{tab:goodness_of_fit_table} presents the $R^2$ and RMSE values for
all test cases. The $R^2$ values are above 99$\%$, indicating that the model
reproduces the test cases with high accuracy. The low RMSE values (mostly below
0.5 mV)  also confirm the validity of the predictions, suggesting that the model
effectively captures the trends in a variety of scenarios.

\begin{table}[h!]
	\centering
	\begin{tabular}{| c | c | c | c | c |}
		\hline

		\multirow{2}{*}{Test case} & \multicolumn{2}{c|}{40 Pa} &
		\multicolumn{2}{c|}{60 Pa}
		\\ \cline{2-5}

		                           & $R^2$ ($\%$)               & RMSE (mV) &
		$R^2$ ($\%$)               & RMSE  (mV)                               \\
		\hline
		1                          & 99.73                      & 0.393     &
		99.79                      & 0.434                                    \\
		\hline
		2                          & 99.85                      & 0.194     &
		99.78                      & 0.302                                    \\
		\hline
		3                          & 99.87                      & 0.284     &
		99.94                      & 0.383                                    \\
		\hline
	\end{tabular}

	\caption{Coefficient of Determination ($R^2$) and Root Mean Squared Error (RMSE) for the test cases.}
	\label{tab:goodness_of_fit_table}

\end{table}

\subsection{Implementation of the potential model in dust dynamics simulation.}
\label{Implementation_in_dust_dynamics_simulation}

We test the utility of the potential model in a simple particle dynamics
simulation. The simulation consists of eight dust grains with radius $R_d=1.69$
$\mu m$ and mass $m_d=3.05\times10^{-14} \, kg$, matching those used in the
DRIAD simulations presented in Subsections \ref{Numerical modeling} and
\ref{Test cases}. For simplicity, we assume a uniform charge for all dust grains, corresponding to the charge obtained from the single-grain DRIAD simulation at each pressure ($Q_d=1760e^-$ for 40 Pa and $Q_d=2007e^-$ for 60 Pa). As shown in Fig. \ref{fig:dust_charge}, dust charge depends on the interparticle distance, an effect that will be addressed in future work. The equation of motion of the dust grains is:

\begin{equation}
	\label{eq:eq_of_motion_dust}
	m_d\ddot{\vec{r}}_d = \vec{F}_d + \vec{F}_E + \vec{F}_n + \vec{F}_B
\end{equation}

The dust-dust interaction $\vec{F}_{d}$ is described using the potential model
Eq. (\ref{fig:potential_example}) with the coefficients in Table
\ref{tab:coefficients_table}:

\begin{equation}
	\label{eq:dust_dust_force}
	\vec{F}_d=-Q_d\vec{\nabla}\Phi.
\end{equation}

The force $\vec{F}_E=Q_d\vec{E}_z$ is due to the interaction of the charged dust
grain with the axial dc switching electric field $E_z$, which matches the
time-averaged field of the ionization waves used in the DRIAD simulations with
$f=500 \, \text{Hz}$ (see Table \ref{tab:sim_params_table}). No external electric
field was applied for confinement.

The neutral drag force $\vec{F}_n=-\beta\vec{v}_d$, where
$\beta=\frac{4}{3}\delta p R_d^2\sqrt{8\pi m_n/(k_BT_n)}$, accounts for the
friction due to the collisions of neutral gas atoms on the dust surface and is
proportional the dust velocity $\vec{v}_d$. The coefficient $\delta$ represents
how the gas atoms are reflected from the dust surface and takes a value of 1.44
for MF spheres \cite{melzer2019physics}. The molecular mass of the neutral gas
(neon) $m_n$ is $3.35\times10^{-26}$ $kg$ and the temperature $T_g$ is $300 \,
	K$.

Finally, the Brownian force $\vec{F}_B=\sqrt{2\beta k_B T_g/\Delta
		t_d}\vec{\xi}$ represents the random kicks exerted
on the dust grains by neutral particles \cite{matthews2020dust}. The dust time
step used is $\Delta t_d=10^{-4} s$ and $\vec{\xi}$ is a vector of normally
distributed random numbers between -1 and 1.

\begin{figure}[ht!]

	\centering
	\includegraphics[width=170mm]{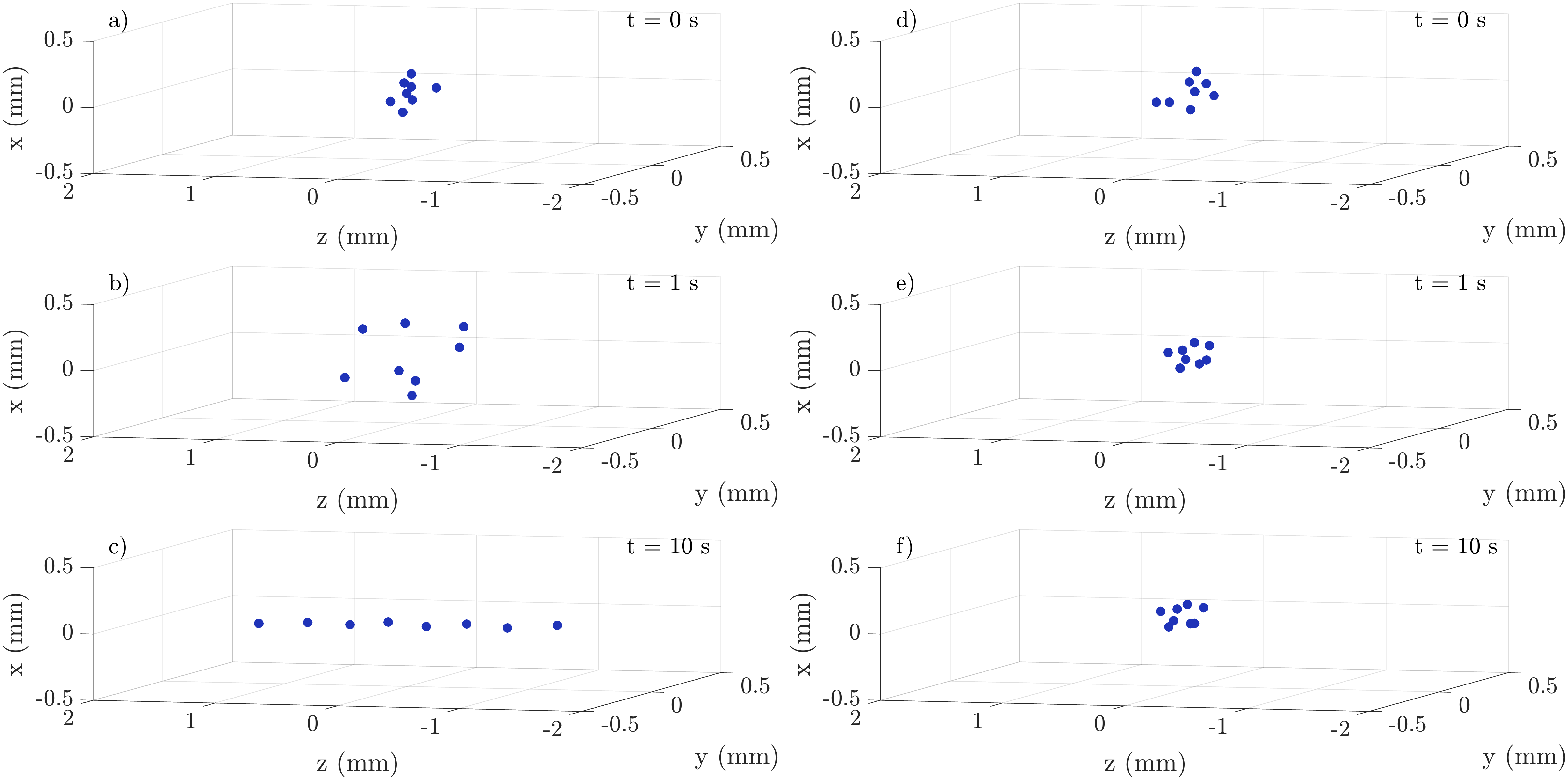}
	\caption{Snapshots of the simulations for: a-c) 40 Pa and d-f) 60 Pa neutral
		gas pressures, showing the temporal evolution of the system at: a, d)
		t=0 s; b, e) t=1 s; and c, f) t=10 s.}
	\label{fig:3D_sim}

\end{figure}

This is a simplified model, since the dust charge is assumed to be fixed, while
simulations with DRIAD showed that dust charge varies by about $5\%$ for dust
particles in close proximity (see Fig. \ref{fig:dust_charge}). The total time
simulated is $10s$ ($10^5$ dust time steps). Fig. \ref{fig:3D_sim} shows snapshots
of the simulations at $t=0 \, s$ (initial positions), $t=1 \, s$ and $t=10 \, s$
(final positions) under 40 Pa and 60 Pa gas pressures. Initial positions of dust
particles are randomly distributed (see Fig. \ref{fig:3D_sim}a, d). In the 40 Pa
case, the dust cloud expands (Fig. \ref{fig:3D_sim}b) while dust grains remain
clustered for 60 Pa (Fig. \ref{fig:3D_sim}e). In the final time step, dust particles
exhibit the formation of a string-like structure in the 40 Pa case (Fig.
\ref{fig:3D_sim}c), with an average interparticle distance of $\sim350$ $\mu m$. In
the 60 Pa case (Fig. \ref{fig:3D_sim}f), the dust cloud does not form an ordered
structure, and dust particles remain fixed with random vibrations around their
equilibrium positions.

The different behavior of the dust cloud in the two cases can be attributed to
the intensity of the axial electric field in the ionization waves (see Fig.
\ref{fig:plasma_params_Ez_switching}). Although the values of $E_z$ for both
pressures are comparable between ionization waves, the magnitudes of the peaks
of $E_z$ in the 40 Pa case ($\sim$2000 V/m) are approximately twice that for 60
Pa ($\sim$1000 V/m) (see Fig. \ref{fig:plasma_params_Ez_switching}). In Ref.
\cite{ivlev2010}, the isotropic-to-string phase transition in dusty plasmas was
studied experimentally and numerically for different electric fields under
microgravity conditions, finding that 6.8 $\mu m$-diameter grains under 10 Pa
gas pressure form string-like structures at electric fields of about 1800 V/m.
In addition, higher electric fields lead to more ordered dust filaments. The
dust cloud returned to the isotropic state when the electric field was reduced.
The same trend was observed in simulations of dust in the PK-4 environment
performed in Ref. \cite{Matthews2021}.

\begin{figure}[ht!]

	\centering
	\includegraphics[width=65mm]{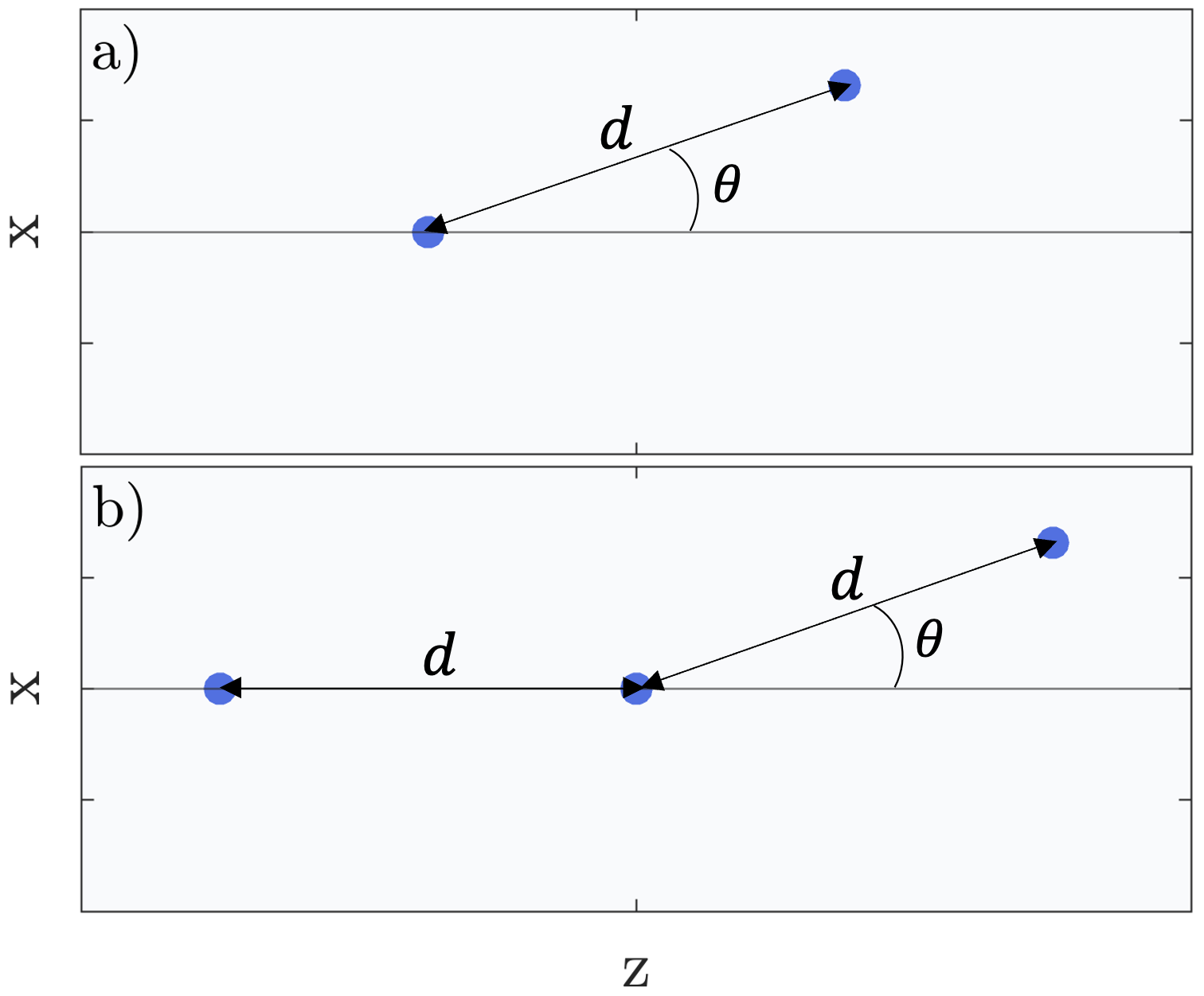}
	\caption{Schematic representation of the dust configurations used for the
		interaction energy distribution analysis: a) two-particle system; b)
		three-particle system.}
	\label{fig:energy_analysis_cases}

\end{figure}

To test this behavior, we analyzed the interaction energy for configurations of two and three dust particles to find configurations that minimize the potential energy.
The total interaction energy of a system of N
charged particles is $U_{int}=(1/2)\sum_i^N\sum_{j\neq
		i}^NQ_i\Phi_{j}(\vec{r}_i)$, where $Q_i$ is the charge of the \textit{i}-th dust
grain and $\Phi_{j}(\vec{r}_i)$ is the electric potential generated by particle
\textit{j} at the position of particle \textit{i}. For $\Phi$, we use the potential model given in Eq.(\ref{eq:potential_model_simplified}) with the corresponding coefficients determined by the machine learning model (Table \ref{tab:coefficients_table}).  For two dust grains, we vary the interparticle distance  ($d$) and
the angle ($\theta$) of their connecting axis with respect to the ion flow
direction (Fig. \ref{fig:energy_analysis_cases}a).  For three dust grains, we fix two particles along the z-axis separated at a distance $\textit{d}$. The third particle is located at the same distance $\textit{d}$ from the second particle, but at an angle $\theta$ relative to the z-axis, as shown in Fig. \ref{fig:energy_analysis_cases}b. If the system is prone to form a filament, the total interaction energy should naturally be minimized when all three particles are collinear, exhibiting a clear minimum near $\theta = 0^\circ$.

\begin{figure}[ht!]

	\centering
	\includegraphics[width=80mm]{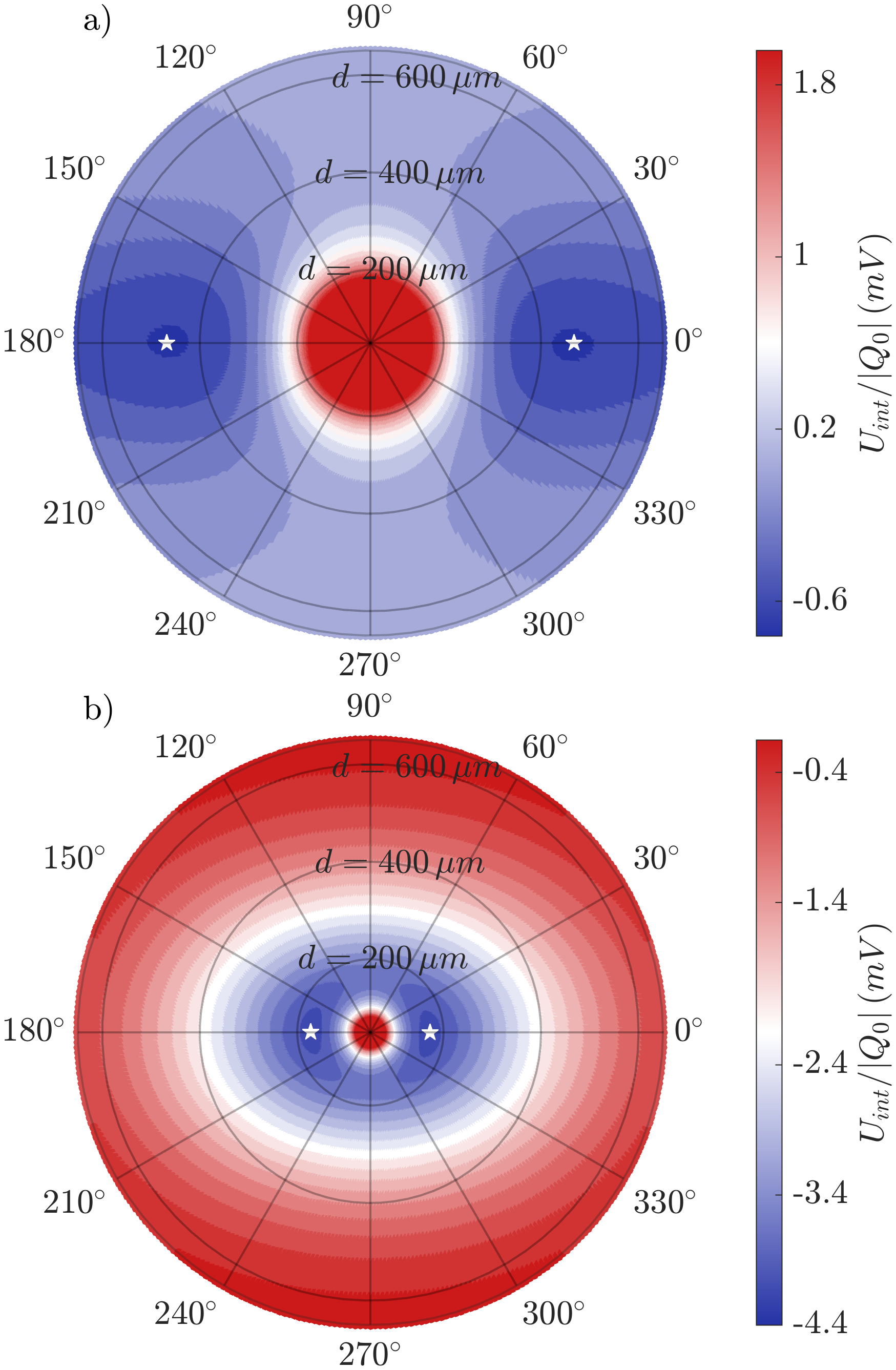}
	\caption{Interaction energy of a system of two dust grains governed by the
		potential model Eq. (\ref{eq:potential_model_simplified}) as a function
		of their separation distance (\textit{d}) and the angle ($\theta$)
		between their connecting axis and the direction of the ion flow (see Fig. \ref{fig:energy_analysis_cases}a), at: a)
		40 Pa and b) 60 Pa. All dust grains are assumed to have charge $Q_0$.
		Data normalized by $|Q_0|$. The positions of the energy minima are represented by white stars.}
	\label{fig:interaction_energy_two}

\end{figure}

\begin{figure}[ht!]

	\centering
	\includegraphics[width=80mm]{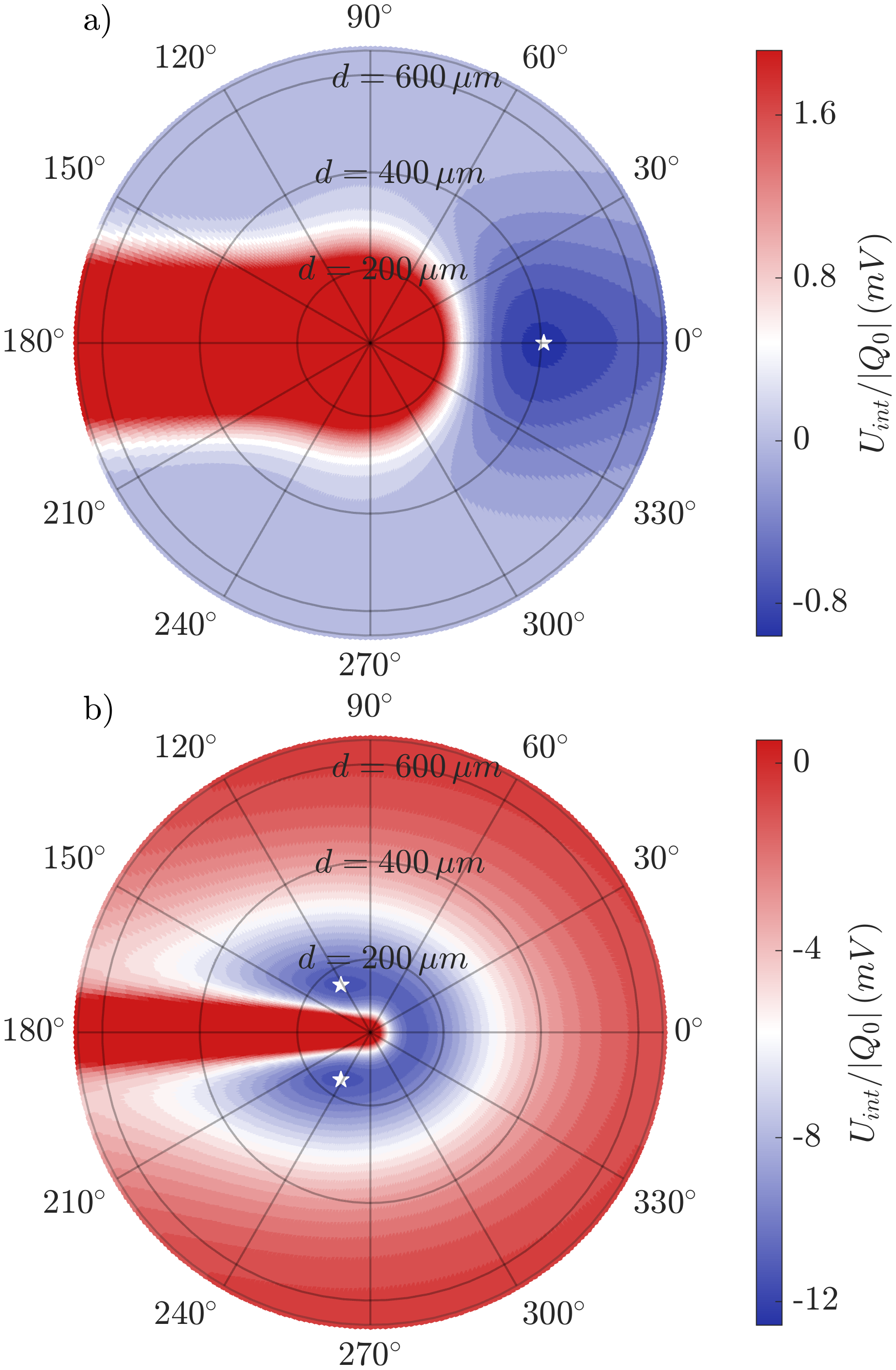}
	\caption{Interaction energy of a system of three dust grains governed by the
		potential model Eq. (\ref{eq:potential_model_simplified}) as a function
		of the interparticle distance (\textit{d}) and the angle ($\theta$) of
		the third particle relative to the z-axis (see Fig. \ref{fig:energy_analysis_cases}b), at: a) 40 Pa and b) 60 Pa. All
		dust grains are assumed to have charge $Q_0$. Data normalized by
		$|Q_0|$. The positions of the energy minima are represented by white stars.}
	\label{fig:interaction_energy_three}

\end{figure}

Fig.~\ref{fig:interaction_energy_two} shows the normalized interaction energy of a
system of two dust grains as a function of the interparticle distance ($d$) and
the angle ($\theta$). In the 40 Pa case (Fig.
\ref{fig:interaction_energy_two}a), two regions of minimum energy appear at
$\theta = 0^\circ$ and $\theta = 180^\circ$ at a distance $d_m
	\approx 460 \, \mu m$. At 60 Pa (Fig. \ref{fig:interaction_energy_two}b), the
interaction energy distribution also exhibits two energy minima near
$\theta=0^\circ$ and $\theta=180^\circ$, but at a shorter interparticle distance
$d_m \approx 175 \, \mu m$. This indicates that for a two-particles system, the
preferred orientation of the dust grains is parallel to the direction of the ion
flow. Fig. \ref{fig:interaction_energy_three} shows the interaction energy of the three-dust-grain system as a function of the particle positions. As shown in Fig. \ref{fig:interaction_energy_three}a, the
system at 40 Pa achieves its minimum energy configuration when $\theta \approx
	0^\circ$, which means that all three particles are aligned with the ion flow forming a
string-like structure. In contrast, the minima for the 60
Pa case (Fig. \ref{fig:interaction_energy_three}b) occur near
$\theta=122^\circ$ and $\theta=238^\circ$. Hence, the equilibrium structure for this scenario is not
a filament, but a triangular lattice.

\section{Conclusions.}

We have used machine learning to build a model that describes the electric
potential of dust and associated ion wakes in the PK-4 experiment. The model was
constructed based on the analysis of key features of the potential distribution
in the vicinity of dust chains, ensuring that the parameters in the model
reflect those properties.

Based on the analysis of the simulation data, the model based on a fully independent formulation
of the shielded dust and Gaussian wake centered on each dust grain (Eq.
(\ref{eq:potential_model})) was simplified, yielding a compact formulation with
only a few coefficients. The simplified model enhances the model's generalizability, minimizes
the risk of overfitting, and improves computational efficiency. The coefficients that characterize the wake shielding,
$\lambda$, $\alpha$, $\sigma_x$ and $\sigma_z$, were obtained from a few
representative linear dust configurations with varying interparticle distances. The resulting
expression for the interaction potential, Eq. \ref{eq:potential_model_simplified}, accurately reproduces the potential distributions obtained from numerical simulations of the ion dynamics, capturing the essential features of the ion wake in a dc switching electric field.

The model was validated by showing that the same set of coefficients
successfully reproduces the electric potential for diverse dust configurations.
This confirms that the parameters retain their validity in different geometries,
demonstrating the model's generalizability and predictive capability. The training and test cases yield $R^2$ values exceeding 99$\%$,
demonstrating the model's high accuracy in different scenarios.

We used the resulting expression for the wake-mediated dust potential to model
the dust dynamics without modeling the ions directly. This approach is
efficient, as it reduces both the complexity of the code and computational cost
compared to explicitly modeling the dust and ions. A simplified simulation
of eight dust grains was performed, resulting in a chain-like structure in the
40 Pa case and a compact cluster in the 60 Pa case (Fig.~\ref{fig:3D_sim}). The difference in the dynamical behavior of the dust at the two pressures was explained in terms of interaction energy. In the 40 Pa case, the interaction energy is minimized when the particles are aligned with the ion flow, favoring the formation of string-like structures. In contrast, in the 60 Pa case, the energy minimum is shifted, leading to a non-filamentary equilibrium configuration (Fig. \ref{fig:interaction_energy_three}). This difference in behavior is mainly due to the greater magnitude of the electric field associated with the ionization waves in the 40 Pa case. This underscores the importance of the nature of the ionization
waves for a given set of plasma conditions in the formation of strings and the
ER behavior.

This method can also be applied to analyze the dust interaction potential in other dusty plasma experiments.  The model used here is applicable for alternating electric fields which produces a symmetric ion wake on both sides of a dust grain.  In unidirectional electric fields, such as in the plasma sheath in a capacitively coupled plasma cell, the wake is only on one side of
the dust grain \cite{vermillion2024interacting}, and the present model must be adapted to account for the arising
asymmetries.

One limitation of the simplified dust dynamics model presented in Subsection \ref{Implementation_in_dust_dynamics_simulation} is that it neglected the variations of the
dust charge in proximity to other grains (Fig. \ref{fig:dust_charge}). This variation will be taken into account in future work, in which we plan to model large populations of dust grains for various
PK-4 plasma conditions to help explain the experimentally observed structures.

\section{Acknowledgements.}

The authors gratefully acknowledge support from the US Department of Energy,
Office of Science, Office of Fusion Energy Sciences under award number
DE-SC-0021334 and the National Science Foundation grant PHY 2308743. Peter Hartmann gratefully acknowledges the hardware support of the NVIDIA Corporation under the NVIDIA GPU Grant Program.

\section{References.}

\bibliography{bibliography}

@book{melzer2019physics,
  author    = {Melzer, Andr{\'e} and others},
  publisher = {Springer},
  title     = {Physics of Dusty Plasmas},
  volume    = {962},
  year      = {2019}
}

@article{thomas1994plasma,
  author    = {Thomas, H and Morfill, GE and Demmel, V and Goree, J and Feuerbacher, B and M{\"o}hlmann, D},
  journal   = {Physical Review Letters},
  number    = {5},
  pages     = {652},
  publisher = {APS},
  title     = {Plasma crystal: Coulomb crystallization in a dusty plasma},
  volume    = {73},
  year      = {1994}
}

@article{chu1994direct,
  author    = {Chu, JH and Lin, I},
  journal   = {Physical Review Letters},
  number    = {25},
  pages     = {4009},
  publisher = {APS},
  title     = {Direct observation of Coulomb crystals and liquids in strongly coupled rf dusty plasmas},
  volume    = {72},
  year      = {1994}
}

@article{pieper1996experimental,
  author    = {Pieper, JB and Goree, J and Quinn, RA},
  journal   = {Journal of Vacuum Science \& Technology A: Vacuum, Surfaces, and Films},
  number    = {2},
  pages     = {519--524},
  publisher = {American Vacuum Society},
  title     = {Experimental studies of two-dimensional and three-dimensional structure in a crystallized dusty plasma},
  volume    = {14},
  year      = {1996}
}

@article{ivlev2008,
  author    = {Ivlev, AV and Morfill, GE and Thomas, HM and R{\"a}th, C and Joyce, G and Huber, P and Kompaneets, R and Fortov, VE and Lipaev, AM and Molotkov, VI and others},
  journal   = {Physical Review Letters},
  number    = {9},
  pages     = {095003},
  publisher = {APS},
  title     = {First observation of electrorheological plasmas},
  volume    = {100},
  year      = {2008}
}

@article{ivlev2010,
  author    = {Ivlev, Alexei V and Brandt, Philip C and Morfill, Gregor E and Rath, Christoph and Thomas, Hubertus M and Joyce, Glenn and Fortov, Vladimir E and Lipaev, Andrey M and Molotkov, Vladimir I and Petrov, Oleg F},
  journal   = {IEEE Transactions on Plasma Science},
  number    = {4},
  pages     = {733--740},
  publisher = {IEEE},
  title     = {Electrorheological complex plasmas},
  volume    = {38},
  year      = {2010}
}

@phdthesis{kompaneets2007complex,
  author = {Kompaneets, Roman},
  school = {LMU},
  title  = {Complex plasmas: Interaction potentials and non-Hamiltonian dynamics},
  year   = {2007}
}

@article{mitic2013three,
  author    = {Mitic, S and Klumov, BA and Khrapak, SA and Morfill, GE},
  journal   = {Physics of Plasmas},
  number    = {4},
  publisher = {AIP Publishing},
  title     = {Three-dimensional complex plasma structures in a combined radio frequency and direct current discharge},
  volume    = {20},
  year      = {2013}
}

@article{pustylnik2020three,
  author    = {Pustylnik, MY and Klumov, Boris and Rubin-Zuzic, Milenko and Lipaev, AM and Nosenko, Volodymyr and Erdle, Daniel and Usachev, AD and Zobnin, AV and Molotkov, VI and Joyce, Glenn and others},
  journal   = {Physical Review Research},
  number    = {3},
  pages     = {033314},
  publisher = {APS},
  title     = {Three-dimensional structure of a string-fluid complex plasma},
  volume    = {2},
  year      = {2020}
}

@article{melzer1996structure,
  author    = {Melzer, A and Schweigert, VA and Schweigert, IV and Homann, A and Peters, S and Piel, A},
  journal   = {Physical Review E},
  number    = {1},
  pages     = {R46},
  publisher = {APS},
  title     = {Structure and stability of the plasma crystal},
  volume    = {54},
  year      = {1996}
}

@article{couedel2018full,
  author    = {Cou{\"e}del, L{\'e}na{\"\i}c and Nosenko, Volodymyr and Rubin-Zuzic, Milenko and Zhdanov, Sergey and Elskens, Y and Hall, T and Ivlev, AV},
  journal   = {Physical Review E},
  number    = {4},
  pages     = {043206},
  publisher = {APS},
  title     = {Full melting of a two-dimensional complex plasma crystal triggered by localized pulsed laser heating},
  volume    = {97},
  year      = {2018}
}

@article{couedel2011wave,
  author    = {Cou{\"e}del, L and Zhdanov, SK and Ivlev, AV and Nosenko, V and Thomas, HM and Morfill, GE},
  journal   = {Physics of Plasmas},
  number    = {8},
  publisher = {AIP Publishing},
  title     = {Wave mode coupling due to plasma wakes in two-dimensional plasma crystals: In-depth view},
  volume    = {18},
  year      = {2011}
}

@article{melzer1999transition,
  author    = {Melzer, A and Schweigert, VA and Piel, A},
  journal   = {Physical Review Letters},
  number    = {16},
  pages     = {3194},
  publisher = {APS},
  title     = {Transition from attractive to repulsive forces between dust molecules in a plasma sheath},
  volume    = {83},
  year      = {1999}
}

@article{schweigert1996alignment,
  author    = {Schweigert, VA and Schweigert, IV and Melzer, A and Homann, A and Piel, A},
  journal   = {Physical Review E},
  number    = {4},
  pages     = {4155},
  publisher = {APS},
  title     = {Alignment and instability of dust crystals in plasmas},
  volume    = {54},
  year      = {1996}
}

@article{qiao2013mode,
  author    = {Qiao, Ke and Kong, Jie and Oeveren, Eric Van and Matthews, Lorin S and Hyde, Truell W},
  journal   = {Physical Review E—Statistical, Nonlinear, and Soft Matter Physics},
  number    = {4},
  pages     = {043103},
  publisher = {APS},
  title     = {Mode couplings and resonance instabilities in dust clusters},
  volume    = {88},
  year      = {2013}
}

@article{melzer2014nonequilibrium,
  author    = {Melzer, Andre and Schella, Andre and Mulsow, Matthias},
  journal   = {Physical Review E},
  number    = {1},
  pages     = {013109},
  publisher = {APS},
  title     = {Nonequilibrium finite dust clusters: Connecting normal modes and wakefields},
  volume    = {89},
  year      = {2014}
}

@article{ishihara2000effect,
  author    = {Ishihara, Osamu and Vladimirov, SV and Cramer, NF},
  journal   = {Physical Review E},
  number    = {6},
  pages     = {7246},
  publisher = {APS},
  title     = {Effect of a dipole moment on the wake potential of a dust grain in a flowing plasma},
  volume    = {61},
  year      = {2000}
}

@article{rocker2012mode,
  author    = {R{\"o}cker, TB and Ivlev, AV and Kompaneets, R and Morfill, GE},
  journal   = {Physics of Plasmas},
  number    = {3},
  publisher = {AIP Publishing},
  title     = {Mode coupling in two-dimensional plasma crystals: Role of the wake model},
  volume    = {19},
  year      = {2012}
}

@article{vermillion2024interacting,
  author    = {Vermillion, Katrina and Banka, Rahul and Mendoza, Alexandria and Wyatt, Bryant and Matthews, L and Hyde, Truell},
  journal   = {Physics of Plasmas},
  number    = {7},
  publisher = {AIP Publishing},
  title     = {Interacting dust grains in complex plasmas: Ion wake formation and the electric potential},
  volume    = {31},
  year      = {2024}
}

@article{Mendoza2025,
  author    = {Mendoza,  A. and Jiménez Martí,  D. and Matthews,  L. S. and Rodríguez Saenz,  B. and Hartmann,  P. and Kostadinova,  E. and Rosenberg,  M. and Hyde,  T. W.},
  doi       = {10.1063/5.0241139},
  issn      = {1089-7674},
  journal   = {Physics of Plasmas},
  number    = {2},
  publisher = {AIP Publishing},
  title     = {Ion density waves driving the formation of filamentary dust structures},
  url       = {http://dx.doi.org/10.1063/5.0241139},
  volume    = {32},
  year      = {2025}
}

@article{Matthews2021,
  author    = {Matthews,  L.S. and Vermillion,  K. and Hartmann,  P. and Rosenberg,  M. and Rostami,  S. and Kostadinova,  E.G. and Hyde,  T.W. and Pustylnik,  M.Y. and Lipaev,  A.M. and Usachev,  A.D. and Zobnin,  A.V. and Thoma,  M.H. and Petrov,  O.F. and Thomas,  H.M. and Novitskiy,  O.V.},
  doi       = {10.1017/s0022377821001215},
  issn      = {1469-7807},
  journal   = {Journal of Plasma Physics},
  number    = {6},
  publisher = {Cambridge University Press (CUP)},
  title     = {Effect of ionization waves on dust chain formation in a DC discharge},
  url       = {http://dx.doi.org/10.1017/S0022377821001215},
  volume    = {87},
  year      = {2021}
}

@article{Hutchinson2002,
  author    = {Hutchinson,  I H},
  doi       = {10.1088/0741-3335/44/9/313},
  issn      = {0741-3335},
  journal   = {Plasma Physics and Controlled Fusion},
  month     = aug,
  number    = {9},
  pages     = {1953–1977},
  publisher = {IOP Publishing},
  title     = {Ion collection by a sphere in a flowing plasma: I. Quasineutral},
  url       = {http://dx.doi.org/10.1088/0741-3335/44/9/313},
  volume    = {44},
  year      = {2002}
}

@book{Brunton2021,
  author    = {Steven L. Brunton, J. Nathan Kutz},
  publisher = {Cambridge University Press},
  title     = {Data-Driven Science and Engineering. Machine Learning, Dynamical Systems, and Control},
  year      = {2002}
}

@article{fortov2005project,
  author    = {Fortov, V and Morfill, G and Petrov, O and Thoma, M and Usachev, A and Hoefner, H and Zobnin, A and Kretschmer, M and Ratynskaia, S and Fink, M and others},
  journal   = {Plasma Physics and Controlled Fusion},
  number    = {12B},
  pages     = {B537},
  publisher = {IOP Publishing},
  title     = {The project ‘Plasmakristall-4’(PK-4)—A new stage in investigations of dusty plasmas under microgravity conditions: First results and future plans},
  volume    = {47},
  year      = {2005}
}

@article{pustylnik2016plasmakristall,
  author    = {Pustylnik, MY and Fink, MA and Nosenko, V and Antonova, Tetyana and Hagl, T and Thomas, HM and Zobnin, AV and Lipaev, AM and Usachev, AD and Molotkov, VI and others},
  journal   = {Review of Scientific Instruments},
  number    = {9},
  publisher = {AIP Publishing},
  title     = {Plasmakristall-4: New complex (dusty) plasma laboratory on board the International Space Station},
  volume    = {87},
  year      = {2016}
}

@article{fortov2005complex,
  author    = {Fortov, VE and Ivlev, AV and Khrapak, SA and Khrapak, AG and Morfill, GE},
  journal   = {Physics Reports},
  number    = {1-2},
  pages     = {1--103},
  publisher = {Elsevier},
  title     = {Complex (dusty) plasmas: Current status, open issues, perspectives},
  volume    = {421},
  year      = {2005}
}

@article{hartmann2020ionization,
  author    = {Hartmann, Peter and Rosenberg, Marlene and Juhasz, Zoltan and Matthews, Lorin S and Sanford, Dustin L and Vermillion, Katrina and Carmona-Reyes, Jorge and Hyde, Truell W},
  journal   = {Plasma Sources Science and Technology},
  number    = {11},
  pages     = {115014},
  publisher = {IOP Publishing},
  title     = {Ionization waves in the PK-4 direct current neon discharge},
  volume    = {29},
  year      = {2020}
}

@article{matthews2020dust,
  author    = {Matthews, Lorin Swint and Sanford, Dustin L and Kostadinova, Evdokiya G and Ashrafi, Khandaker Sharmin and Guay, Evelyn and Hyde, Truell W},
  journal   = {Physics of Plasmas},
  number    = {2},
  publisher = {AIP Publishing},
  title     = {Dust charging in dynamic ion wakes},
  volume    = {27},
  year      = {2020}
}

@article{gehr2025structural,
  author  = {Gehr, Emerson and Terrell, Abigail and Vermillion, Katrina and Mendoza, Alexandria and Andrew, Bradley and Jimenez Marti, Diana and Kostadinova, Evdokiya and Hartmann, Peter and Matthews, Lorin and Hyde, Truell},
  journal = {arXiv preprint arXiv:2505.14576},
  title   = {Structural States of Filamentary Microgravity Dusty Plasma},
  year    = {2025}
}

@misc{mathworksoverfitting,
  author = {MathWorks},
  title  = {Overfitting},
  url    = {https://www.mathworks.com/discovery/overfitting.html},
  year   = {2025}
}

@misc{mathworksRsquare,
  author = {MathWorks},
  title  = {Coefficient of determination (R-squared)},
  url    = {https://www.mathworks.com/help/stats/coefficient-of-determination-r-squared.html},
  year   = {2025}
}

\end{document}